\documentclass[12pt, notitlepage,aps,preprintnumbers,superscriptaddress,nofootinbib,aps,prd,floatfix]{revtex4-2}
\usepackage{physics}
\usepackage{bm}                
\usepackage{graphicx}
\usepackage{epstopdf}          
\usepackage{hyperref}
\usepackage{multirow}
\usepackage{array}
\newcolumntype{C}[1]{>{\centering\let\newline\\\arraybackslash\hspace{0pt}}m{#1}}
\usepackage{xcolor}
\usepackage[draft]{todonotes}
\usepackage{wrapfig}
\usepackage{ifthen}                 
\usepackage{pdfcomment}             
\usepackage{adjustbox}              
\usepackage{listings}
\usepackage{version}
\usepackage{amssymb}
\usepackage{booktabs}
\usepackage{makecell}
\usepackage{enumitem}
\setlist{nolistsep}
\definecolor{nicered}{rgb}{0.5,0.,0.}
\definecolor{nicegreen}{rgb}{0.,0.5,0.}
\definecolor{niceblue}{rgb}{0.,0.,0.5}
\hypersetup{colorlinks,citecolor=nicegreen,linkcolor=nicered,urlcolor=niceblue}
\usepackage{orcidlink}

\newcommand{\GeV}{\textrm{GeV}}
\newcommand{\HT}{$\hat{H}_{T}$ }       
\newcommand{\HTot}{$\hat{H}_{T}/2$ }   
\newcommand{\HTxt}{$2\hat{H}_{T}$ }     
\newcommand{\ptj}{$p_{T}^{\rm jet}$ }     

\newcommand{\ptjxt}{$2p_{T}^{\rm jet}$ }

\newcommand{\mjj}{$m_{12}$ }
\newcommand{\mjjot}{$m_{12}/2$ }
\newcommand{\mjjxt}{$2m_{12}$ }

\newcommand{\ptmax}{$p_{T,1}e^{0.3y^*}$ }

\usepackage[normalem]{ulem}

\begin{document}

\author{Alim Ablat}
\affiliation{School of Physics Science and Technology, Xinjiang University, Urumqi, Xinjiang 830046 China\looseness=-1}
\author{Sayipjamal Dulat\,\orcidlink{0000-0003-2087-0727}}
\email{sdulat@hotmail.com}
\affiliation{School of Physics Science and Technology, Xinjiang University, Urumqi, Xinjiang 830046 China\looseness=-1}
\affiliation{Department of Physics and Astronomy, Michigan State University, East Lansing, MI 48824, USA\looseness=-1}
\author{Tie-Jiun Hou}
\affiliation{School of Nuclear Science and Technology, University of South China, Hengyang, Hunan 421001, China\looseness=-1}
\author{Joey Huston\,\orcidlink{0000-0001-9097-3014}}
\affiliation{Department of Physics and Astronomy, Michigan State University, East Lansing, MI 48824, USA\looseness=-1}
\author{Pavel Nadolsky\,\orcidlink{0000-0003-3732-0860}}
\affiliation{Department of Physics and Astronomy, Michigan State University, East Lansing, MI 48824, USA\looseness=-1}
\affiliation{Department of Physics, Southern Methodist University, Dallas, TX 75275-0181, USA}
\author{Ibrahim Sitiwaldi}
\affiliation{School of Physics Science and Technology, Xinjiang University, Urumqi, Xinjiang 830046 China\looseness=-1}
\author{Keping Xie\,\orcidlink{0000-0003-4261-3393}}
\affiliation{Department of Physics and Astronomy, Michigan State University, East Lansing, MI 48824, USA\looseness=-1}
\author{C.-P. Yuan\,\orcidlink{0000-0003-3988-5048}}
\affiliation{Department of Physics and Astronomy, Michigan State University, East Lansing, MI 48824, USA\looseness=-1}

\collaboration{CTEQ-TEA Collaboration}
\preprint{MSUHEP-24-020}

\title{
The impact of LHC precision measurements of inclusive jet and dijet production on the CTEQ-TEA global PDF fit}

\begin{abstract}

In this study, we investigate the impact of new LHC inclusive jet and dijet measurements on parton distribution functions (PDFs) that describe the proton structure, with a particular focus on the gluon distribution at large momentum fraction, $x$, and the corresponding partonic luminosities. We assess constraints from these datasets using next-to-next-to-leading-order (NNLO) theoretical predictions, accounting for a range of uncertainties from scale dependence and numerical integration. From the scale choices available for the calculations, our analysis shows that the central predictions for inclusive jet production show a smaller scale dependence than dijet production.
We examine the relative constraints on the gluon distribution provided by the inclusive jet and dijet distributions, and also explore the phenomenological implications for inclusive $H$, $t\bar{t}$, and $t\bar{t}H$ production at the  LHC at 14 TeV.
\end{abstract}

\maketitle

\vspace{-10pt}
\tableofcontents

\section{Introduction}

Global parton distribution function (PDF) fits have been carried out by several groups using data from both fixed target and collider experiments, utilizing a wide range of processes, and covering a wide kinematic range. Current PDF fits, such as the CT18 next-to-next-to-leading order (NNLO) PDFs\cite{Hou:2019efy}, contain over 4000 data points.  Most of the recent data added to the fits from the LHC. Among the most crucial processes used in the global PDF fits is jet production, which provides sensitivity to PDFs, and especially to the gluon distribution, over a wide range of parton $x$ values. Precision jet data thus play a key role in global PDF determination efforts, complementing constraints from processes such as Drell-Yan \cite{ATLAS:2018pyl,ATLAS:2019fgb,ATLAS:2017rue,CMS:2019raw,LHCb:2016zpq,LHCb:2021huf} and $t\bar{t}$ \cite{ATLAS:2019hxz,ATLAS:2023gsl,ATLAS:2020ccu,CMS:2018adi,CMS:2021vhb} production at the LHC. Extensive studies on the impact of Drell-Yan and $t\bar t$ data on CTEQ-TEA PDFs were presented in Refs. \cite{Sitiwaldi:2023jjp,Ablat:2023tiy}.

\begin{figure}[tb]
 	\includegraphics[width=0.70\linewidth]{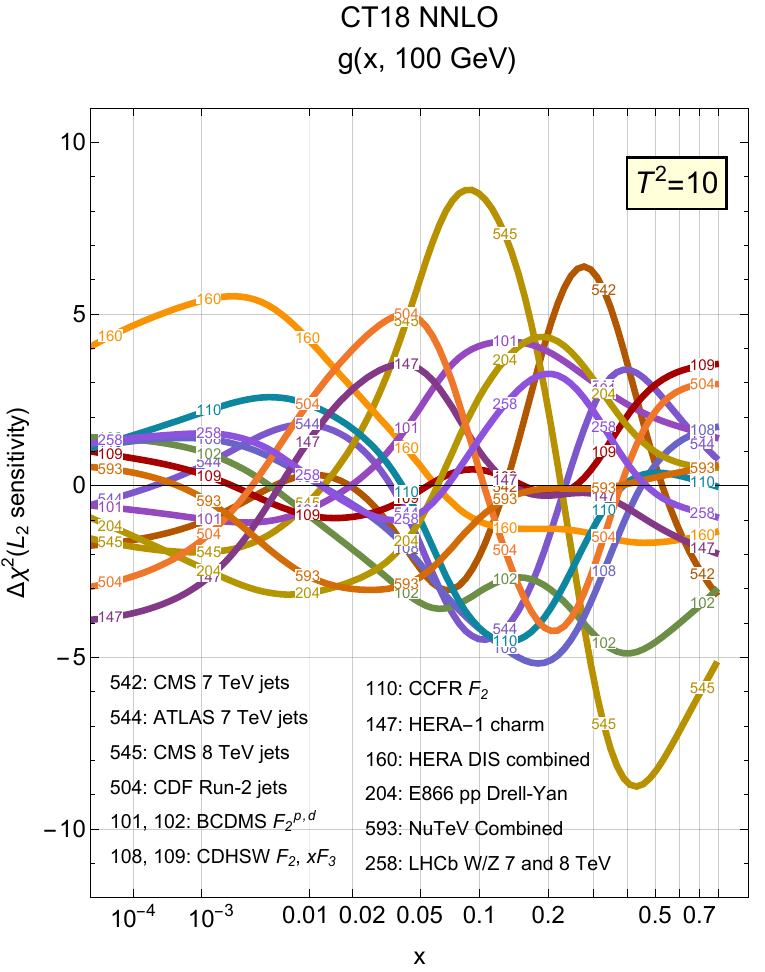}
	\caption{Sensitivities of various experiments in the CT18 NNLO analysis obtained using the $L_2$ sensitivity method \cite{Jing:2023isu}. The curves illustrate the approximate changes in $\chi^2$ for specified data sets upon the increase of $g(x,Q)$ at specified $x$ and $Q=100$ GeV by one standard deviation corresponding to the increase in 10 units in the total $\chi^2$. }
	\label{fig:L2CT18}
 \end{figure}

Jet production has been an important component of global PDF fits since the measurements carried out at the Tevatron~\cite{CDF:2008hmn,D0:2008nou,D0:2009rxw,D0:2010mik} and has continued with the inclusive jet measurements at center-of-mass energies of $\sqrt{s} = 2.76$ TeV \cite{ATLAS:2013pbc,CMS:2015jdl}, $\sqrt{s} = 7$ TeV \cite{ATLAS:2014riz,CMS:2014nvq}, $\sqrt{s} = 8$ TeV \cite{ATLAS:2017kux,CMS:2016lna}, and $\sqrt{s} = 13$ TeV \cite{ATLAS:2017ble,CMS:2016jip,CMS:2021yzl}  as well as dijet measurements at $\sqrt{s} = 7$ TeV \cite{ATLAS:2013jmu,CMS:2012ftr}, $\sqrt{s} = 8$ TeV \cite{CMS:2017jfq}, and $\sqrt{s} = 13$ TeV \cite{ATLAS:2017ble,CMS:2023fix} from the ATLAS and CMS collaborations at the LHC.
One of the most straightforward ways to display the impact of the different data sets in a global PDF fit is through the use of the $L_2$ sensitivity. The $L_2$ sensitivity indicates the pull that a particular data set has on a particular PDF (or PDF luminosity).  The $L_2$ sensitivity for the gluon distribution for the CT18 NNLO PDFs is shown in Fig.~\ref{fig:L2CT18}, with only the experiments having the greatest impact on the gluon being shown. The impact of the jet production data can be seen over the entire parton $x$ range. It is also evident that the individual jet data sets often pull the gluon distributions in different directions at any particular $x$ value. 

Although PDF fits have been carried out at leading order and next-to-leading order, the most precise results are obtained at NNLO. By definition, this requires the use of NNLO predictions, which are available at both leading color~\cite{Currie:2017eqf,Currie:2016bfm,Gehrmann-DeRidder:2019ibf,Currie:2018xkj,Liu:2018ktv} and full color~\cite{Czakon:2019tmo,Chen:2022tpk}. Jet production measurements can be performed for inclusive jets (where every jet above the transverse momentum threshold contributes to the cross-section), or for dijets, most often using the two leading transverse momentum jets, but also possible using all combinations of jets above the threshold. Inclusive jet production, as its name implies, has the advantage of being completely inclusive, a strength from both the experimental and theoretical perspectives. Dijet production includes more information on the event kinematics. In this paper, we explore the constraints on PDFs provided by both types of measurements. Note that potentially both cross-sections could be used at the same time in a global fit if the statistical correlations are known. Currently, these are known only to the LHC experiments but may be released in the future. An additional complication is that the inclusive jets and dijet measurements are often conducted separately, with different naming conventions for systematic errors. 

Jets at the LHC are most commonly measured with the anti-$k_T$  jet algorithm~\cite{Cacciari:2008gp}. Jet production is almost unique for analyses at the LHC, in that most measurements are carried out for two jet sizes, rather than a single one,  $R$=0.4 or $R$=0.6 (for ATLAS), and $R$=0.4\footnote{Note that CMS used $R$=0.5 for the smaller jet size in Run 1.} and $R$=0.7 (for CMS). The jet shape is better described at NNLO than at NLO, fixed order predictions tend to be more accurate at NNLO at larger values of $R$. Thus, where available, we use the larger $R$ size\footnote{Better agreement with the experimental jet shape can be obtained by the use of parton shower Monte Carlo programs which make use of matrix element information at NLO. However, as the accuracy is only NLO, they are not sufficient to use for PDF fits at NNLO.}. 

A calculation of jet production at NNLO is quite CPU-intensive. The codes are complex enough that the calculations have been carried out by the theory collaborations themselves, and the results are made available in grids that can be used in the determination of jet cross-sections. 
Recently, Ref.~\cite{Britzger:2022lbf} presented NNLO calculations, based on the leading-colour and leading $N_f$ approximations, of inclusive jet 
and dijet 
cross sections for jet measurements~\cite{CMS:2015jdl,ATLAS:2014riz,CMS:2014nvq,ATLAS:2017kux,CMS:2016lna,ATLAS:2017kux,ATLAS:2017ble,CMS:2016jip,CMS:2021yzl,ATLAS:2013jmu,CMS:2012ftr,CMS:2017jfq}. 
The inclusive jet cross section was calculated using two different choices for the renormalization scale ($\mu_R$) and the factorization scale ($\mu_F$): \HT (representing the scalar sum of the transverse momenta of all partons) and \ptj (denoting the transverse momentum of each jet). In contrast, the dijet cross-section calculations utilized a common scale $\mu_R=\mu_F=m_{12}$. For the CMS 8 TeV dijet data, in addition to calculations using the scale $\mu_R=\mu_F=m_{12}$, calculations employing the scale \ptmax were also provided.
These results have been implemented in the \texttt{APPLfast} interpolation grids~\cite{Britzger:2022lbf}, a NNLO descendant of \texttt{APPLgrid}~\cite{Carli:2010rw} and \texttt{fastNLO}~\cite{Kluge:2006xs,Britzger:2012bs}, which are publicly available on the Ploughshare repository~\cite{PloughshareProject}.
The grids often have fluctuations that are of the same size as the precision desired, so a smoothing operation may need to be carried out, possibly along with the use of additional statistical uncertainty, to achieve a $\chi^2$ value for the data set that indicates reasonable agreement with the theory. There are several ways of carrying out this smoothing/additional statistical uncertainty, and it is crucial to determine whether the procedure adopted affects the resultant PDF determination. In the past, conclusions have been drawn regarding the level of agreement between data and theory, for example between inclusive jet and dijet production, that have been impacted by the level of precision of the corresponding grids.

There is also more ambiguity about the optimum scale choice for dijet production than for inclusive jet production. For inclusive jet production, typically a scale choice related to transverse momentum \ptj has been used, appropriate given the inclusive nature of the measurement. The NNLOJET~\cite{Gehrmann:2018szu,Currie:2018oxh} collaboration has shown that an equivalent level of prediction can be obtained with the use of a scale \HT, where  \HT  is the sum of all transverse momenta jets in the event above a given $p_T$ threshold. A scale of the leading jet transverse momentum $p_T^{j_1}$, formerly used, leads to some undesirable behavior at low transverse momentum~\cite{Currie:2018xkj}.

For dijets, the natural scale choice would be the dijet mass. There is the ambiguity mentioned above as to whether only the two leading jets should be used, or all jets above threshold. Another caveat for dijet measurements is that the character of the hard interaction for a given value of $m_{12}$ is different in the cases where both jets are central or, for example, the jets are separated by a large rapidity range, or some other combination. A double, or even triple, differential measurement, for example in $m_{12}$ and $y^*$ can remove some of this ambiguity when used with a scale such as \ptmax. In this paper, we will examine the consequences of the choice of scale for both inclusive jets and dijets. 

The detail of the experimental measurements of jet production can change with the region of the detector in which they are carried out, and thus the jets are often calibrated as a function of rapidity. The shifts in the nuisance parameters of the measurements can often be different among the different rapidity regions, and decorrelation models for those systematic errors must be determined by the experimental collaborations if a reasonable value of $\chi^2$ is to be achieved. It is also important to understand whether any bias in the PDF determination results from the imposition of those decorrelations. All these factors will also be considered in this paper. 

This paper is organized as follows:
In Sec.~\ref{Sec:Overview}, we begin by reviewing jet measurements and then analyze the impact of decorrelation of the systematic errors in the ATLAS 8 TeV and 13 TeV inclusive jet datasets on the central values and uncertainty bands of the gluon PDF.
In Sec.~\ref{Sec:EpumpStudy}, 
we discuss theoretical predictions and examine the scale dependence of the gluon PDF using the \texttt{ePump} package (Error PDF Updating Method Package)~\cite{Schmidt:2018hvu,Hou:2019gfw}, which enables the rapid determination of both the central PDF and the uncertainties of the updated PDFs.
In Sec.~\ref{Sec:Globalfit},
we incorporate the optimal inclusive jet data in the global analysis within the CT18 framework. 
We present the impact of each new dataset on the gluon PDF by adding them one by one, and simultaneously, on top of the CT18 baseline. Furthermore, we provide the gluon-gluon ($gg$) luminosity with its uncertainty and discuss the phenomenological implications for the production of inclusive Higgs bosons, top-quark pairs, and associated $t\bar{t}H$ production at the LHC 14 TeV.
Sec.~\ref{Sec:Conclusion} contains our conclusion and outlook.

\section{Inclusive jet and dijet measurements at the LHC}\label{Sec:Overview}

 The ATLAS and CMS collaborations have conducted numerous measurements of single-inclusive and dijet cross sections at various center-of-mass energies, ranging from $\sqrt{s} = 2.76$ TeV to 13 TeV. All inclusive jet measurements—at $\sqrt{s} = 2.76$ TeV \cite{ATLAS:2013pbc,CMS:2015jdl}, $\sqrt{s} = 7$ TeV \cite{ATLAS:2014riz,CMS:2014nvq}, $\sqrt{s} = 8$ TeV \cite{ATLAS:2017kux,CMS:2016lna}, and $\sqrt{s} = 13$ TeV \cite{ATLAS:2017ble,CMS:2016jip,CMS:2021yzl}—are presented as functions of the jet transverse momentum ($p_T$) and absolute jet rapidity ($|y|$). However, dijet measurements are presented either double-differentially or triple-differentially.

In the double-differential case, ATLAS measurements at 7 TeV \cite{ATLAS:2013jmu} and 13 TeV \cite{ATLAS:2017ble} are shown as a function of the dijet invariant mass ($m_{12}$) and half the rapidity separation ($y^*$) of the two leading jets. The half rapidity separation is defined as \( y^* = |y_1 - y_2|/2 \), where $y_1$ and $y_2$ represent the rapidities of the leading and subleading jets, respectively. Similarly, CMS measurements at 7 TeV \cite{CMS:2012ftr} are presented as a function of the dijet invariant mass ($m_{12}$) and the largest absolute rapidity ($|y_{\max}|$) of the two leading jets. Here, $y_{\max}$ is defined as
\[
y_{\max} = \textrm{sign}\left( |\max(y_1, y_2)| - |\min(y_1, y_2)| \right) \cdot \max(|y_1|, |y_2|).
\]

In the triple-differential case, the CMS measurement at 8 TeV \cite{CMS:2017jfq} is presented as a function of the average transverse momentum, $p_{T,\textrm{avg}}$
\footnote{$p_{T,\textrm{avg}} =(p_{T,1} + p_{T,2})/2$, where $p_{T,1}$ and $p_{T,2}$ represent the transverse momenta of the leading and subleading jets, respectively.}, half the rapidity separation $y^*$, and the boost $y^b$ of the two leading jets, where $y^b =|y_1 + y_2|/2$. In contrast, the CMS measurement at 13 TeV \cite{CMS:2023fix} is presented as a function of $y^b$, $y^*$, and $m_{12}$.

\renewcommand{\arraystretch}{1.5}
\begin{table}[!h]
\caption{Summary of inclusive jet and dijet measurements analyzed in this study.
For each dataset, we list the experiment, reference, center-of-mass energy, integrated luminosity ($\mathcal{L}_{\rm int}$), number of data points ($N_{\rm pt}$), jet radius ($R$), observable, covered kinematic region, and the renormalization and factorization scales. 
	}
    \begin{tabular}{lccccccccc}
        \toprule
        Expt.  & Ref.                  & \makecell{$\sqrt{s}$\\ \textrm{[TeV]}} & \makecell{$\mathcal{L}_{\rm int}$ \\ $[\textrm{fb}^{-1}]$ }& $N_{\rm pt}$ &\makecell{Jet radius\\  $R$} & Observable &   \makecell{Coverage\\ $[p_T$ \textrm{ in GeV}]}  & \makecell{Central\\ $\mu_R,\ \mu_F$} & Decor.  \\
        \hline
        \multicolumn{10}{c}{Inclusive jet data sets in CT18 \cite{Hou:2019efy}} \\
        \hline
       	ATLAS & \cite{ATLAS:2014riz} & 7               & 4.5                & 140      & 0.4,0.6  & $\frac{\dd^2\sigma}{\dd p_T\dd|y|}$ &  \makecell{ $ |y| < 3.0 $ \\ $p_T^{\rm jet}\in [100,1992]$}    & $p_T^{\rm jet}$ & Yes\\
        CMS   & \cite{CMS:2014nvq}   & 7               & 5.0                & 158      & 0.5,0.7  & $\frac{\dd^2\sigma}{\dd p_T\dd|y|}$ &  \makecell{ $ |y| < 3.0 $ \\ $p_T^{\rm jet}\in [114,2116]$}   & $p_T^{\rm jet}$ & No\\
        CMS   & \cite{CMS:2016lna}   & 8               & 19.7               & 185      & 0.7  & $\frac{\dd^2\sigma}{\dd p_T\dd|y|}$ &  \makecell{ $ |y| < 4.7 $ \\ $p_T^{\rm jet}\in [21,2500]$}   & $p_T^{\rm jet}$ &No\\
        \hline
        \multicolumn{10}{c}{New inclusive jet data} \\
        \hline 
        ATLAS & \cite{ATLAS:2017kux} & 8               & 20.3               & 171      & 0.4,0.6    & $\frac{\dd^2\sigma}{\dd p_T\dd|y|}$ &  \makecell{ $ |y| < 3.0 $ \\ $p_T^{\rm jet}\in [70,2500]$}             & $p_T^{\rm jet}\mbox{ or }\hat{H}_T$          &  Yes \\
        ATLAS & \cite{ATLAS:2017ble} & 13              & 3.2                & 177      & 0.4    & $\frac{\dd^2\sigma}{\dd p_T\dd|y|}$ &  \makecell{ $ |y| < 3.0 $ \\ $p_T^{\rm jet}\in [100,3937]$}            & $p_T^{\rm jet} \mbox{ or } \hat{H}_T  $       & Yes \\
        CMS   & \cite{CMS:2021yzl}   & 13              & 33.5               & 78       & 0.4,0.7    & $\frac{\dd^2\sigma}{\dd p_T\dd|y|}$ &  \makecell{ $ |y| < 2.0 $ \\ $p_T^{\rm jet}\in [97,3103]$}             & $p_T^{\rm jet} \mbox{ or } \hat{H}_T $        &  No \\
        \hline
        \multicolumn{10}{c}{New dijet data} \\
        \hline
        ATLAS & \cite{ATLAS:2013jmu} & 7               & 4.5                & 90        & 0.4,0.6     & $\frac{\dd^2\sigma}{\dd m_{12}\dd y^*}$         & \makecell{$ y^* < 3.0 $     \\ $m_{12}\in[260,5040]$}                         & $m_{12}$                & No                              \\
        CMS   & \cite{CMS:2012ftr}   & 7               & 5.0                & 54        & 0.7     & $\frac{\dd^2\sigma}{\dd m_{12}\dd|y_{\max}|}$   & \makecell{$ y_{\max} < 2.5 $ \\ $m_{12}\in[197,5058]$ }                        & $m_{12}$               & No                              \\
        CMS   & \cite{CMS:2017jfq}   & 8               & 19.7               & 122       & 0.7     & $\frac{\dd^3\sigma}{\dd p_{T,\rm avg}\dd y^*\dd y^b}$  & \makecell{$ y^{*}(y^{b}) < 3.0$ \\ $p_{T,\rm avg}\in[133,1784]$}        & \makecell{$m_{12}$ \mbox{ or } \\ $p_{T,1} e^{0.3 y*}$ } & No        \\
        ATLAS & \cite{ATLAS:2017ble} & 13              & 3.2                & 136       & 0.4     & $\frac{\dd^2\sigma}{\dd m_{12}\dd y^*}$         & \makecell{$ y^* < 3.0 $     \\ $m_{12}\in[260,9066]$}                         & $m_{12}$                              & No                \\
        \bottomrule
    \end{tabular}    
    \label{tab:DataInThisWork}
\end{table}

Table~\ref{tab:DataInThisWork} presents the inclusive jet and dijet datasets analyzed in this study.
In all experiments, jets are reconstructed using the anti-$k_T$ algorithm~\cite{Cacciari:2008gp} with varying jet radii $R$~\cite{Ellis:1993tq,Dokshitzer:1997in,Cacciari:2008gp}.
This study excludes jet measurements at $\sqrt{s} = 2.76$ TeV from ATLAS with $\mathcal{L}_{\rm int} = 0.2~\textrm{pb}^{-1}$~\cite{ATLAS:2013pbc} and CMS with $\mathcal{L}_{\rm int} = 5~\textrm{pb}^{-1}$~\cite{CMS:2015jdl}, due to their very low integrated luminosities.
Recently, the CMS Collaboration published dijet results at $\sqrt{s} = 13$ TeV~\cite{CMS:2023fix}, which will be incorporated into future work. In this analysis, we use the full set of systematic uncertainties and correlations available from the HEPData repository~\cite{Maguire:2017ypu} for all included measurements.

The ATLAS Collaboration~\cite{ATLAS:2017kux} recommended various decorrelation options (Table 4 in Ref.~\cite{ATLAS:2017kux}) for the systematic uncertainties, \emph{i.e.}, Jet Energy Scale (JES) Flavor Response, JES Multi-Jet Balance Fragmentation, and JES Pile-up Rho Topology. These uncertainties are not fully correlated across all rapidity jet bins.
The recommended decorrelation strategies consist of 18 options: option 1 \dots option 12 split the systematic errors into two sub-components, while option 13 \dots option 18 split them into three sub-components, based on rapidity and jet transverse momentum ($p_T^{\text{jet}}$).
In this work, we apply these decorrelation options to all relevant uncertainties and label them as Option1 \dots Option18, respectively, with the ID matching Table 4  of Ref.~\cite{ATLAS:2017kux}. Additionally, we also check the ATLAS recommendations~\cite{ATLAS:2017kux} for $R=0.4$ and $R=0.6$, respectively, and denote them as SuggestedR4 and SuggestedR6.
SuggestedR4 employs decorrelation option 7 for Jet Energy Scale (JES) Flavor Response, option 17 for JES Multi-Jet Balance Fragmentation, and option 18 for JES Pile-up Rho Topology. Similarly, SuggestedR6 utilizes decorrelation option 7 for JES Flavor Response, option 17 for JES Multi-Jet Balance Fragmentation, and option 16 for JES Pile-up Rho Topology.

\begin{figure}[!h]

	\includegraphics[width=0.49\linewidth]{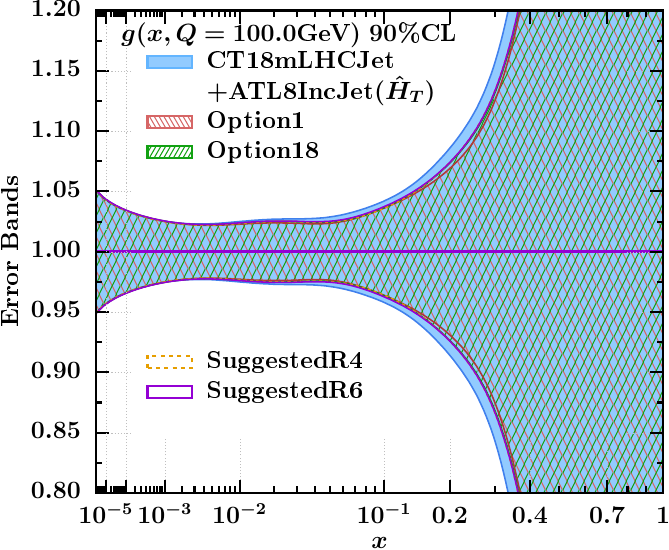}
	\includegraphics[width=0.49\linewidth]{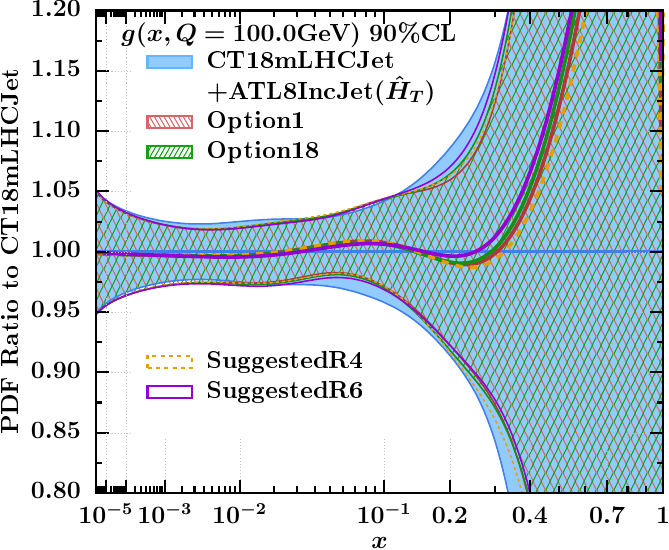}
	\includegraphics[width=0.49\linewidth]{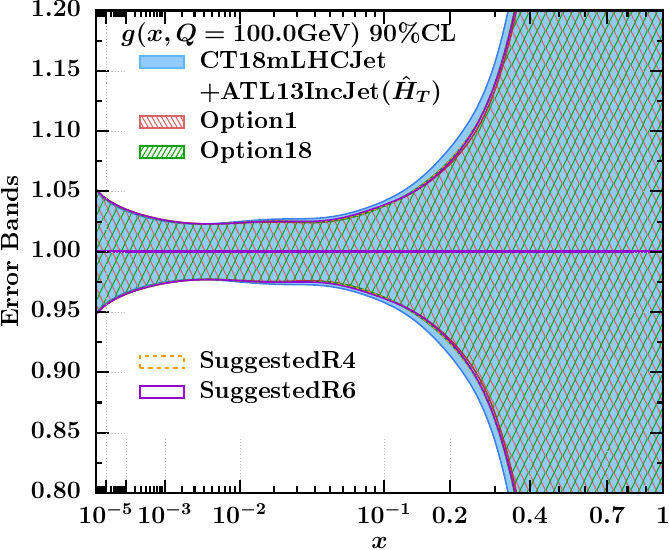}
	\includegraphics[width=0.49\linewidth]{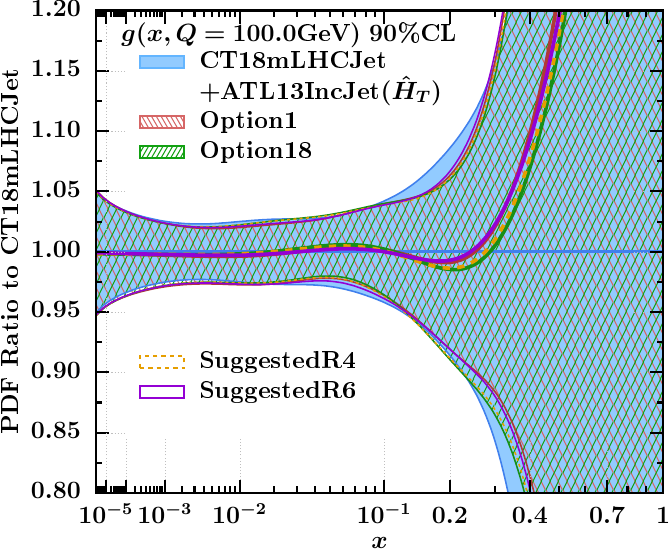}
	\caption{The impact of the ATLAS 8 TeV and 13 TeV inclusive jet datasets on the central value and error band of the gluon PDF at the 90\% confidence level, after decorrelation the relevant systematic errors using \texttt{ePump}. }
	\label{Fig:DecorImpactePump}
\end{figure}

We perform the decorrelation for both ATLAS 8 TeV and 13 TeV inclusive jet datasets.
Using \texttt{ePump} profiling~\cite{Schmidt:2018hvu,Hou:2019gfw},  we present $\chi^2$ from different decorrelation options in Table \ref{tab:DecorChi2ePump},
  and the corresponding PDFs  in Fig.~\ref{Fig:DecorImpactePump} for scales $\hat{H}_T$ and $p_T^{\text{jet}}$, respectively.
We find that the two-component splitting option has a negligible impact on $\chi^2$, whereas the three-component splitting option results in a significant improvement in $\chi^2$ values. For the ATLAS 8 TeV data, SuggestedR6 proves to be a favorable option, while for the ATLAS 13 TeV inclusive jet data, Option18 provides the best $\chi^2$. The impact on the PDF central values and error bands is minimal across all options, consistent with the conclusions reported in ATLASpdf21~\cite{ATLAS:2021vod} and MSHT~\cite{Harland-Lang:2017ytb}. 
In Fig.~\ref{Fig:DecorImpactePump}, we present representative results for option 1 from the two-component splitting method, option 18 from the three-component splitting method, and SuggestedR4(6) options, all of which show similar impacts on the PDF central values and error bands.

In this work, we will utilize the SuggestedR6  for the ATLAS 8 TeV inclusive jet data and adopt option 18 for the ATLAS 13 TeV inclusive jet data, as these options provide the best $\chi^2$ values.

\renewcommand{\arraystretch}{1}
\begin{table}[h]
    \centering
	\caption{Comparison of the $\chi^2/N_{\rm pt}$ from \texttt{ePump} for various decorrelation options of the systematic uncertainties for the ATLAS 8/13 TeV inclusive jet datasets with  $\hat{H}_T$ and $p_T^{\text{jet}}$ scales, respectively.
    }
    \begin{tabular}{lcccc}
        \toprule
        Option & \multicolumn{2}{c}{ATLAS 8 TeV IncJet} & \multicolumn{2}{c}{ATLAS 13 TeV IncJet} \\        \cline{2-5}
        Scale & $\hat{H}_{T}$ & {$p_{T}^{\rm jet}$} & $\hat{H}_{T}$ & {$p_{T}^{\rm jet}$} \\
        \hline
        Fully correlate & 3.04 & 3.07 & 2.87 & 4.10 \\
        Option1 & 3.03 & 3.76 & 2.87 & 4.10 \\
        Option2 & 3.04 & 3.77 & 2.87 & 4.10 \\
        Option3 & 3.04 & 3.77 & 2.87 & 4.10 \\
        Option4 & 3.04 & 3.77 & 2.87 & 4.10 \\
        Option5 & 3.03 & 3.76 & 2.82 & 4.05 \\
        Option6 & 3.03 & 3.76 & 2.86 & 4.09 \\
        Option7 & 2.99 & 3.68 & 2.82 & 4.03 \\
        Option8 & 3.03 & 3.75 & 2.85 & 4.07 \\
        Option9 & 3.04 & 3.77 & 2.87 & 4.10 \\
        Option10 & 3.01 & 3.74 & 2.87 & 4.10 \\
        Option11 & 3.04 & 3.77 & 2.87 & 4.10 \\
        Option12 & 3.03 & 3.76 & 2.87 & 4.10 \\
        Option13 & 2.84 & 3.54 & 2.85 & 4.08 \\
        Option14 & 2.85 & 3.57 & 2.81 & 4.03 \\
        Option15 & 2.94 & 3.66 & 2.85 & 4.08 \\
        Option16 & 2.84 & 3.56 & 2.79 & 4.01 \\
        Option17 & 2.90 & 3.61 & 2.79 & 4.00 \\
        Option18 & 2.83 & 3.52 & 2.76 & 3.98 \\
        SuggestedR4 & 2.76 & 3.45 & 2.78 & 3.99 \\
        SuggestedR6 & 2.75 & 3.47 & 2.79 & 4.01 \\
        \bottomrule
    \end{tabular}
    \label{tab:DecorChi2ePump}
\end{table}

\section{Selection criteria for the global fit}\label{Sec:EpumpStudy}

In this section, we pursue two primary objectives.
First, we describe the theoretical framework adopted in this study (Sec.~\ref{Sec:TheoryTreatment}) and examine the corresponding scale dependence of the inclusive jet and dijet production (Sec. \ref{Sec:ScaleImpact}). 
Second, we assess which of the inclusive jet or dijet datasets, collected at the same center-of-mass energy and integrated luminosity of the LHC using the same detector, offers stronger constraints on the gluon PDF error bands (Sec. \ref{Sec:Opt}). 
The detailed treatment of Monte Carlo fluctuations in the theoretical predictions is also provided (Sec.\ref{Sec:TheoryTreatment}). These analyses are performed through global fitting or by employing various tools that utilize a broad range of statistical procedures based on either Monte Carlo or Hessian methods. Examples include \texttt{ePump} \cite{Schmidt:2018hvu,Hou:2019gfw}, Hessian profiling \cite{Paukkunen:2014zia,xFitter:2022zjb}, Bayesian reweighting techniques \cite{Giele:1998gw,Ball:2011gg,Ball:2010gb}, and the \texttt{PDFSense}/$L_2$-sensitivity method \cite{Wang:2018heo,Jing:2023isu}.
 
As discussed in Sec.~\ref{Sec:Overview} the CT18 NNLO global QCD analyses incorporated 
ATLAS 7 TeV inclusive jet (ATL7IncJet)~\cite{ATLAS:2014riz}, CMS 7 TeV inclusive jet (CMS7IncJet)~\cite{CMS:2014nvq} and CMS 8 TeV inclusive jet (CMS8IncJet)~\cite{CMS:2016lna} measurements.

This study also utilizes dijet data from the ATLAS and CMS experiments, specifically from the 7 and 8 TeV datasets, as well as the ATLAS 13 TeV dataset.\cite{ATLAS:2013jmu,CMS:2012ftr,CMS:2017jfq,ATLAS:2017ble}.
However, since the CT18 baseline already includes some of the counterpart inclusive data sets from these measurements, directly including these dijet datasets would lead to double counting unless the dataset cross-correlations are also implemented. 
To circumvent the issue of double counting and compare the impact of inclusive jet data relative to dijet data, we introduced a new intermediate global data set, CT18mLHCJet, which uses the same framework as the CT18 analysis (such as the treatment of experimental correlated systematic errors and the tolerance criterion for defining the PDF uncertainties) but excludes the inclusive jet datasets (ATLAS 7 TeV \cite{ATLAS:2014riz}, CMS 7 TeV \cite{CMS:2014nvq} and CMS 8 TeV \cite{CMS:2016lna}) already included in CT18. 

Our results in this section are obtained using the \texttt{ePump} package~\cite{Schmidt:2018hvu,Hou:2019gfw}, which enables the extraction of best-fit PDFs and associated PDF uncertainties. Previous studies utilizing \texttt{ePump} can be found in Refs.~
\cite{Willis:2018yln,Hou:2019gfw,Yalkun:2019gah,Czakon:2019yrx,Ablat:2020wzd,Czakon:2019yrx,Kadir:2020yml,Rashidin:2024aji,Yang:2022bxv}.

\subsection{ Theory treatment }\label{Sec:TheoryTreatment}

\begin{figure}[h]

	\includegraphics[width=0.49\linewidth]{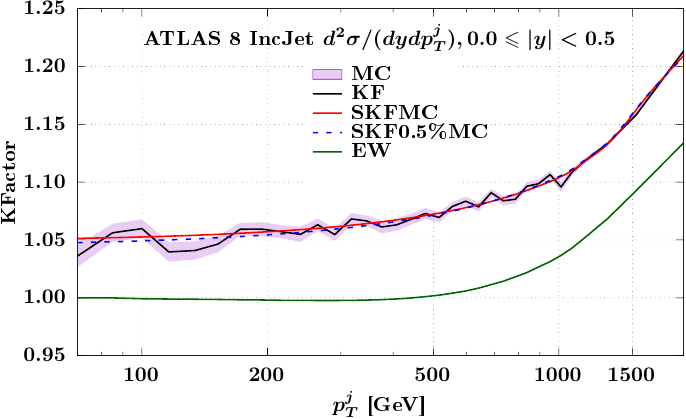}
	\includegraphics[width=0.49\linewidth]{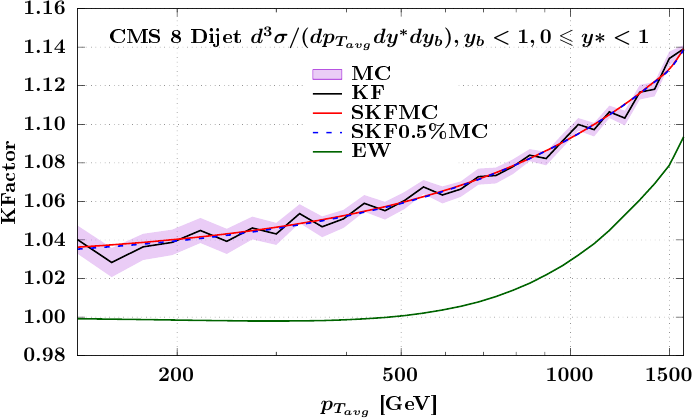}
	\caption{Comparison of different K-factor treatments for ATL8IncJet (left) and CMS8DiJet (right) is presented. The plots display the original K-factor (black line), the MC error from the grid (purple band), the smoothed K-factor using a 0.5\% MC error (SKF0.5\%MC, dashed blue line), and the smoothed K-factor using the MC error from the grid (SKFMC, red line). Additionally, the EW K-factors corresponding to the ATLAS 8 TeV inclusive jet and dijet data are shown (green line).
 }
\label{KF:smoothed}
\end{figure}

In our analysis, we performed PDF fit using \texttt{ePump} \cite{Schmidt:2018hvu,Hou:2019gfw} by including the new inclusive jet and dijet datasets in Table \ref{tab:DataInThisWork} and theoretical predictions at NNLO QCD  obtained by convoluting the \texttt{APPLfast} tables~\cite{Britzger:2022lbf} with the CT18mLHCJet PDFs. In addition, we also applied both non-perturbative and electroweak (EW) corrections, provided by the corresponding experiments\cite{ATLAS:2017kux,ATLAS:2017ble,CMS:2021yzl,ATLAS:2013jmu,CMS:2012ftr,CMS:2017jfq}.
However, we encountered difficulties in achieving a good $\chi^2/N_{\rm pt}$ for our fit, with $\chi^2/N_{\rm pt}= 2.75$, $2.76$, and $2.04$ for the ATLAS 8 TeV (ATL8IncJet), ATLAS 13 TeV (ATL13IncJet), and CMS 13 TeV (CMS13IncJet) inclusive jet datasets, respectively. For the dijet datasets, 
the $\chi^2/N_{\rm pt}= 1.81, 1.93$, $2.76$, and $2.66$ for the ATLAS 7 TeV (ATL7DiJet), CMS 7 TeV (CMS7DiJet), CMS 8 TeV (CMS8DiJet), and ATLAS 13 TeV (ALT13DiJet) datasets, respectively. 
This prompted us to consider potential uncertainties arising from the Monte Carlo (MC) integration in the theoretical calculations. As a representative example, we present the MC errors of the NNLO K-factor (KF), defined as the ratio of theoretical predictions at NNLO to NLO for a given $p_T$, extracted from the grid
\footnote{
The grid was generated using the NNPDF3.1 \cite{NNPDF:2017mvq} set, which includes MC errors. 
We applied the same percentage of MC error across all PDFs in this study.}
for the ATL8IncJet and CMS8DiJet data in Fig. \ref{KF:smoothed}, labeled as MC (purple band). 
We observe that the NNLO K-factor value can fluctuate around a smooth curve by approximately 1\%, with an associated MC error of about 0.5\% for jet $p_T$ below roughly 200 GeV. Although small, this error is non-negligible when compared to the statistical uncertainty in the high-precision data and must therefore be included in the fit.

To minimize the impact of MC statistical fluctuations on the quality of PDF fits, we applied smoothing techniques to the NNLO K-factor. 

A smoother curve for the NNLO K-factor leads to a reduced $\chi^2$ value. However, the best-fit result generally retains some sensitivity to the chosen functional form. This sensitivity can be conservatively estimated by incorporating an uncorrelated MC integration error.

One possible method for incorporating these uncertainties is to treat them as an additional bin-by-bin source of uncorrelated systematic error, similar to the approach used by the NNPDF group \cite{Gehrmann-DeRidder:2016ycd,AbdulKhalek:2020jut}. 
Another approach, employed by the CT18 \cite{Hou:2019efy} and MSHT \cite{Harland-Lang:2017ytb} groups, involves smoothing the NNLO K-factor using a smooth function (SKF). An alternative method, as described in \cite{Carrazza:2017bjw}, uses neural networks to achieve smooth interpolation of the NNLO K-factor.
 We categorize five different treatments of theory, following the methods adopted by CTEQ-TEA \cite{Hou:2019efy}, MSHT \cite{Harland-Lang:2017ytb}, and NNPDF \cite{Gehrmann-DeRidder:2016ycd, AbdulKhalek:2020jut}, as follows:
\begin{description}
    \item [Method 1 -- NNLO]
    Using the original NNLO theory prediction \cite{Britzger:2022lbf} alone.
    \item [Method 2 -- NNLOMC] Using the NNLO theory prediction along with the MC error, and treating the latter as an uncorrelated systematic uncertainty, is an approach adopted by NNPDF~\cite{Gehrmann-DeRidder:2016ycd, AbdulKhalek:2020jut}.
    \item [Method 3 -- SKF] Smooth the NNLO $K$-factor using continuous, regular functions. This method was adopted by
    MSHT \cite{Harland-Lang:2017ytb} group.
    \item [Method 4 -- SKF0.5\%MC] 
    In addition to applying the smooth NNLO $K$-factor, we include an additional overall 0.5\% MC error in the theory prediction, treating it as an uncorrelated systematic uncertainty. This approach was adopted in the CT18 analysis~\cite{Hou:2019efy}.
    \item [Method 5 -- SKFMC]
  In addition to applying the smooth NNLO $K$-factor, we incorporate the MC uncertainty extracted from the grid \cite{Britzger:2022lbf}, treating it as an uncorrelated systematic error. This approach forms the basis of our final results.
\end{description}
Fig. \ref{KF:smoothed} illustrates the representative smoothed $K$-factor values for the ATL8IncJet and CMS8DiJet datasets. We observe that the SKF0.5\%MC method (blue dashed line) and the SKFMC method (red solid line) produce very similar $K$-factor values.

\begin{figure}[!h]

    \includegraphics[width=0.49\linewidth]{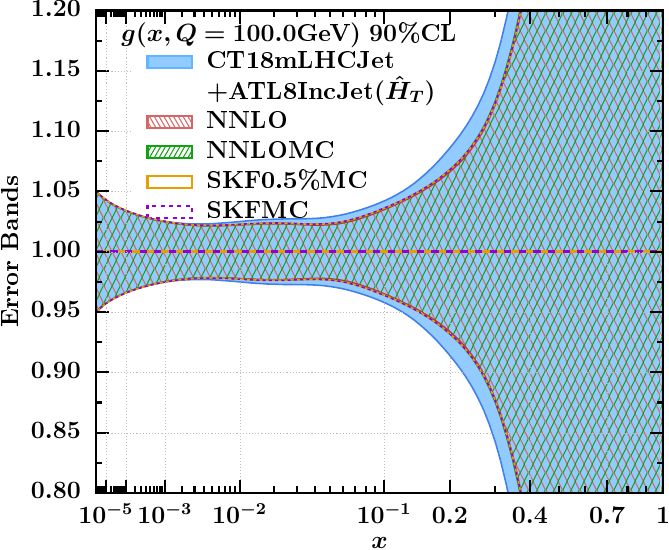}
    \includegraphics[width=0.49\linewidth]{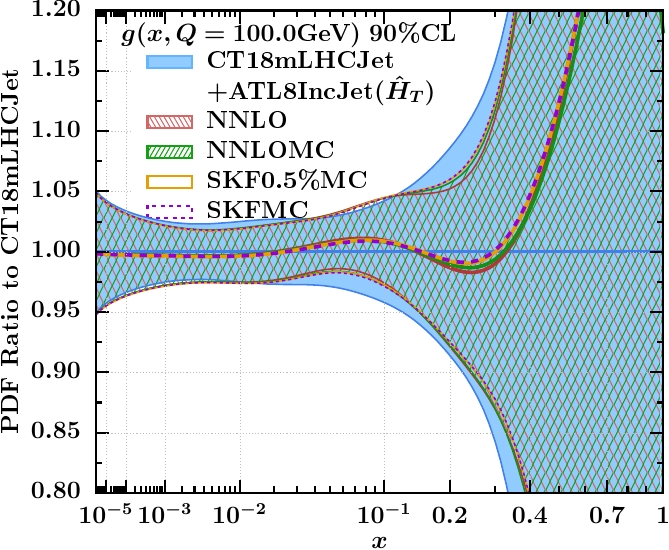}
    \includegraphics[width=0.49\linewidth]{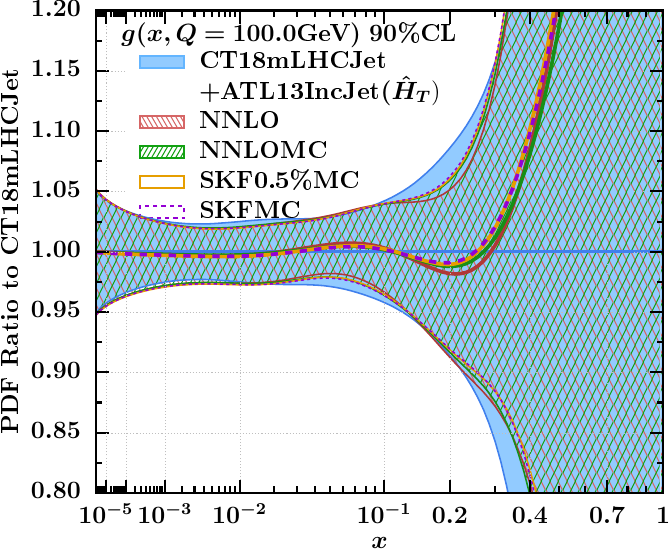}
    \includegraphics[width=0.49\linewidth]{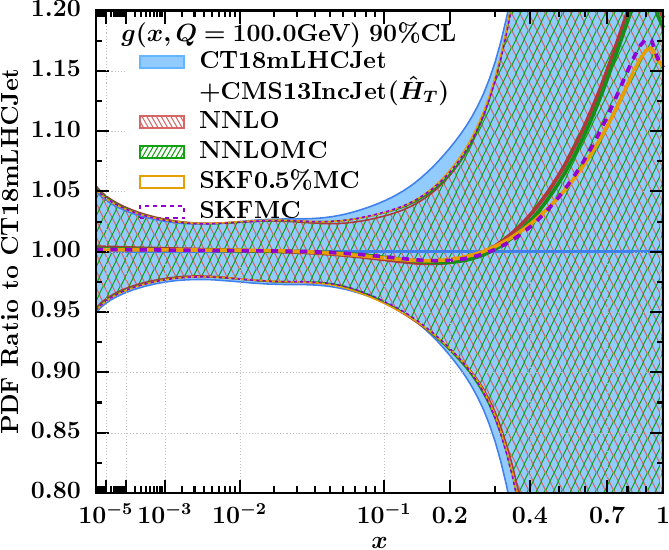}
    \caption{ The upper left panel shows the gluon PDF error band for ATL8IncJet production with various theory treatments. The remaining panels display the ratio of the \texttt{ePump} updated gluon PDF to the CT18mLHCJet gluon PDF. The shaded areas represent the PDF uncertainty at 90\%.C.L.  }
    \label{Fig:DiffThImpactIncJet}
\end{figure}

\begin{figure}[!h]
    \includegraphics[width=0.49\linewidth]{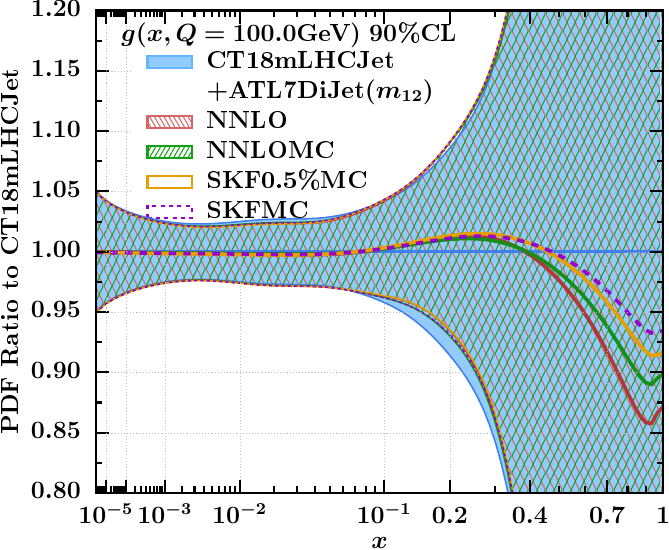}
    \includegraphics[width=0.49\linewidth]{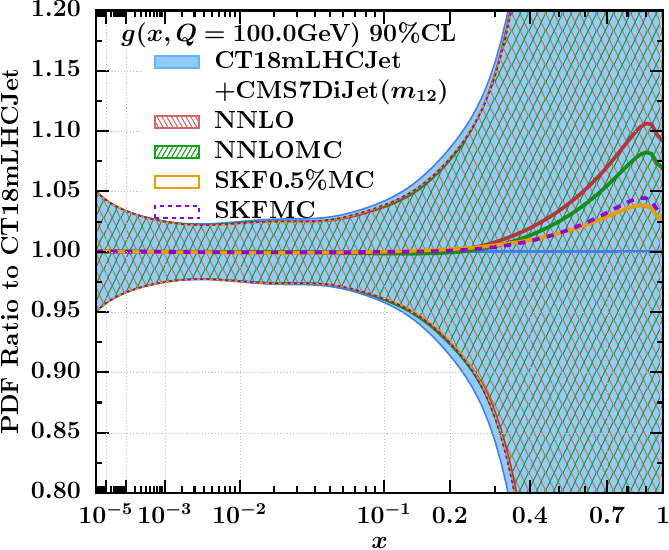}
    \includegraphics[width=0.49\linewidth]{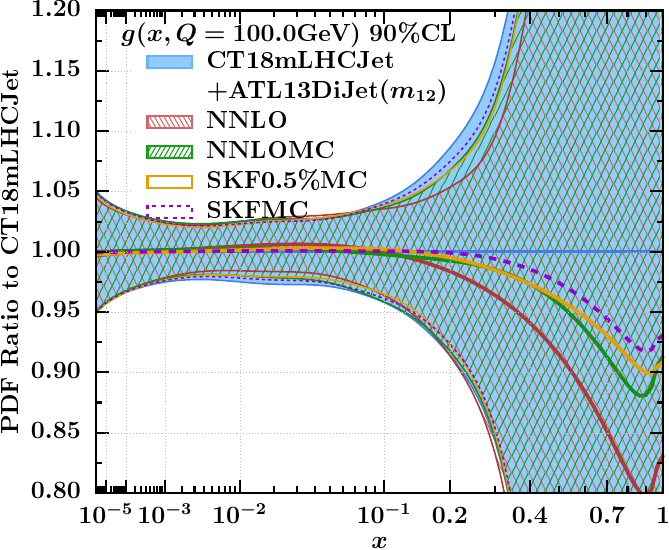}
    \includegraphics[width=0.49\linewidth]{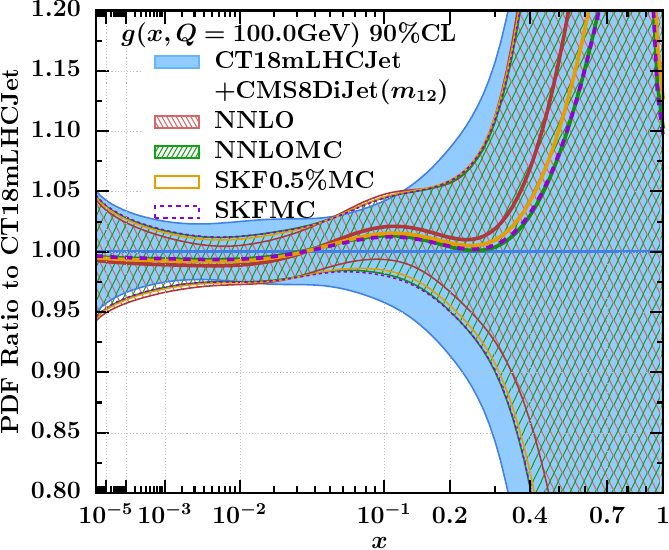}
    \includegraphics[width=0.49\linewidth]{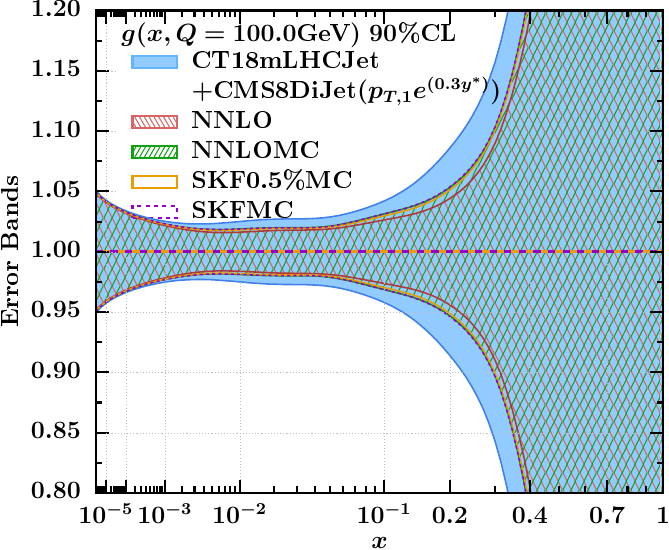}
    \includegraphics[width=0.49\linewidth]{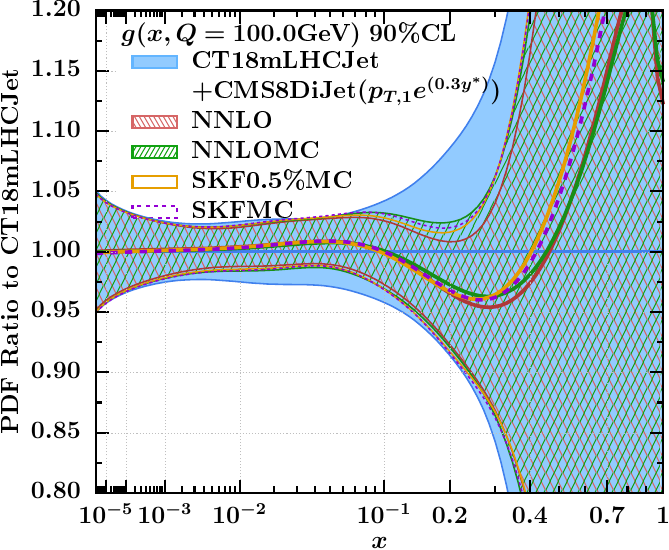}
    \caption{Same as Fig. \ref{Fig:DiffThImpactIncJet}, but for dijet datasets (ATL7DiJet, CMS7DiJet, CMS8DiJet and ATL13DiJet).  }
    \label{Fig:DiffThImpactDijet}
\end{figure}

\renewcommand{\arraystretch}{1}
\begin{table}
\caption{The quality-of-fit in terms of $\chi^2/N_{\rm pt}$.
Comparison of the $\chi^2/N_{\rm pt}$ values from \texttt{ePump} is presented, incorporating the inclusive jet datasets (ATL8IncJet, ATL13IncJet, and CMS13IncJet) sequentially on top of the CT18mLHCJet baseline. The comparison uses different theoretical treatments (NNLO, NNLOMC, SKF, SKFMC, SKFP0.5\%MC) and considers scale choices such as \HT and \ptj, along with factor-of-two variations.}
    \centering
    \begin{tabular}{l |C{1.5cm}C{1.5cm}C{1.5cm}|C{1.5cm}C{1.5cm}C{1.5cm}}
   	\hline
   	Central scale  & \multicolumn{3}{c|}{\HT} & \multicolumn{3}{c}{\ptj}\\                              

    Variation &  1/2 & 1 & 2 & 1/2 & 1 & 2 \\
   	\hline
   	\multicolumn{7}{c}{ATLAS 8 TeV IncJet, $N_{\rm pt}=171$} \\ \hline
        NNLO          & 3.59 & 2.75 & 2.50 & 6.03 & 3.47 & 2.67 \\
        NNLOMC        & 2.45 & 1.99 & 1.89 & 3.84 & 2.34 & 1.91 \\
        SKF            & 1.96 & 2.01 & 2.13 & 2.00 & 1.95 & 2.01 \\
        SKF0.5\%MC     & 1.49 & 1.54 & 1.67 & 1.56 & 1.46 & 1.51 \\        
        SKFMC         & 1.48 & 1.53 & 1.66 & 1.52 & 1.46 & 1.51 \\

   	\hline
   	\multicolumn{7}{c}{ATLAS 13 TeV IncJet, $N_{\rm pt}=177$}\\ 
   	\hline
        NNLO      & 3.97 & 2.76 & 2.30 & 7.59 & 3.98 & 2.72 \\
        NNLOMC    & 2.14 & 1.62 & 1.45 & 3.58 & 2.03 & 1.52 \\
        SKF        & 1.96 & 1.87 & 1.88 & 2.03 & 1.81 & 1.77 \\
        SKF0.5\%MC & 1.45 & 1.40 & 1.44 & 1.54 & 1.35 & 1.32 \\        
        SKFMC     & 1.34 & 1.27 & 1.29 & 1.38 & 1.21 & 1.18 \\

   	\hline
   	\multicolumn{7}{c}{CMS 13 TeV IncJet, $N_{\rm pt}=78$}\\ 
   	\hline
   	NNLO       & 2.57  & 2.04 & 1.87  & 3.92  & 2.47  & 1.98  \\
   	NNLOMC     & 1.71  & 1.43 & 1.37  & 2.48  & 1.64  & 1.37  \\
   	SKF        & 1.71  & 1.63 & 1.64  & 1.55  & 1.58  & 1.57  \\
   	SKF0.5\%MC& 1.13  & 1.09 & 1.14  & 1.01  & 1.02  & 1.02  \\     
   	SKFMC      & 1.20  & 1.16 & 1.20  & 1.08  & 1.09  & 1.10  \\

   	\hline
   \end{tabular}
   \label{tab:Chi2ePumpIncJet}
\end{table}

\begin{table}
	\caption{ Similar to Table \ref{tab:Chi2ePumpIncJet}, but for dijet (ATL7DiJet, CMS7DiJet, CMS8DiJet, and ATL13DiJet) datasets with scale as \mjj or \ptmax scales.}
    \centering
   \begin{tabular}{l |C{1.5cm}C{1.5cm}C{1.5cm}|C{1.5cm}C{1.5cm}C{1.5cm}}
   	\hline
   	Central Scale  & \multicolumn{3}{c|}{\mjj}     & \multicolumn{3}{c}{\ptmax}      \\    
    Variation    &   1/2 & 1 & 2 & 1/2 & 1 & 2 \\
   	\hline
   	\multicolumn{7}{c}{ATLAS 7 TeV dijet, $N_{\rm pt}=90$}                                                                                                    \\ \hline
   	NNLO         & 2.90 & 1.81 & 1.49 &   --   &  --    &  --    \\
   	NNLOMC       & 1.29 & 0.98 & 0.95 &   --   &  --    &  --    \\
   	SKF          & 1.21 & 1.13 & 1.20 &   --   &  --    &  --    \\
   	SKF0.5\%MC  & 1.16 & 1.05 & 1.10 &   --   &  --    &  --   \\    
   	SKFMC        & 0.81 & 0.79 & 0.87 &   --   &  --    &  --    \\

   	\hline
   	\multicolumn{7}{c}{CMS 7 TeV dijet, $N_{\rm pt}=54$}                          \\
   	\hline
   	NNLO         & 2.16 & 1.93 & 1.92 &  --    &  --    &  --    \\
   	NNLOMC       & 1.87 & 1.70 & 1.70 &  --    &  --    &  --    \\
   	SKF          & 1.73 & 1.74 & 1.84 &  --    &  --    &  --    \\
   	SKF0.5\%MC  & 1.57 & 1.60 & 1.70 &  --    &  --    &  --    \\    
   	SKFMC        & 1.55 & 1.55 & 1.63 &  --    &  --    &  --    \\

   	\hline
   	\multicolumn{7}{c}{ATLAS 13 TeV dijet, $N_{\rm pt}=136$}                       \\
   	\hline
   	NNLO         & 4.33 & 2.76 & 2.22 &  --    &  --    &  --    \\
   	NNLOMC       & 1.50 & 1.10 & 0.99 &  --    &  --    &  --    \\
   	SKF          & 1.51 & 1.53 & 1.71 &  --    &  --    &  --    \\
   	SKF0.5\%MC  & 1.24 & 1.23 & 1.35 &  --    &  --    &  --    \\    
   	SKFMC        & 0.90 & 0.87 & 0.93 &  --    &  --    &  --    \\

   	\hline
   	\multicolumn{7}{c}{CMS 8 TeV  dijet, $N_{\rm pt}=122$}                         \\
   	\hline
   	NNLO         & 3.14 & 2.66 & 3.33 & 5.36 & 2.92 & 2.19 \\
   	NNLOMC       & 1.31 & 1.35 & 2.00 & 2.14 & 1.30 & 1.13 \\
   	SKF          & 1.46 & 1.86 & 2.86 & 2.00 & 1.59 & 1.56 \\
   	SKF0.5\%MC  & 1.10 & 1.41 & 2.25 & 1.41 & 1.13 & 1.15 \\    
   	SKFMC        & 0.95 & 1.20 & 1.90 & 1.25 & 1.00 & 1.01 \\

   	\hline
    \end{tabular}
    \label{tab:Chi2ePumpDijet}
\end{table}

We applied the smoothing procedure to all jet datasets listed in Table \ref{tab:DataInThisWork}. Subsequently, we investigated the impact of these different theoretical treatments on the CT18mLHCJet PDFs. Using \texttt{ePump} \cite{Schmidt:2018hvu,Hou:2019gfw}, we incorporated one dataset at a time on the top of the CT18mLHCJet baseline. The corresponding $\chi^2/N_{\rm pt}$ values for each inclusive jet and dijet data are presented in Table \ref{tab:Chi2ePumpIncJet} and Table \ref{tab:Chi2ePumpDijet}. Additionally, the comparison of the corresponding updated gluon PDF at $Q=100~\GeV$ is presented in Fig. \ref{Fig:DiffThImpactIncJet} for inclusive jets and in Fig. \ref{Fig:DiffThImpactDijet} for dijets, respectively.

As shown in Tables \ref{tab:Chi2ePumpIncJet} and \ref{tab:Chi2ePumpDijet}, the $\chi^2/N_{\rm pt}$ values for all datasets significantly improved, reaching acceptable levels for the final fits. For example, the ATL8IncJet dataset, which initially had the largest $\chi^2/N_{\rm pt}$ under the original NNLO theory (\textbf{Method 1}), achieves an acceptable $\chi^2/N_{\rm pt}$ after smoothing the $K$-factors and treating the bin-by-bin MC error as an uncorrelated (\textbf{Method 5}). Similar improvements are observed across other datasets. This approach is designed to reduce fluctuations caused by MC errors and improve the $\chi^2/N_{\rm pt}$ without introducing bias or unexpected impact on the PDFs. However, it is essential to examine its impact on the PDF error bands and best fit values, which are presented in Figs. \ref{Fig:DiffThImpactIncJet} and \ref{Fig:DiffThImpactDijet}. 

The impact of the inclusive jet data is shown in Fig. \ref{Fig:DiffThImpactIncJet}.
The upper-left panel illustrates the impact of the ATL8IncJet dataset on the CT18mLHCJet gluon PDF error bands under various theoretical treatments.
The remaining panels show the ratio of the updated gluon PDFs from the inclusive jet datasets (ATL8IncJet, ATL13IncJet, and CMS13IncJet) for different theoretical treatments. All error bands are presented at the 90\% confidence level (CL). 

We observe that the different theoretical treatments have a similar impact on the gluon PDF error bands (upper-left panel of Fig \ref{Fig:DiffThImpactIncJet} ). 
The changes in the central values are minimal and can be considered negligible. Similar conclusions apply to the ATL13IncJet and CMS13IncJet datasets, therefore the corresponding error band plots are omitted.

In Fig. \ref{Fig:DiffThImpactDijet}, 
we plot the impact of various theoretical treatments on dijet measurements. We observe that the error bands for the ATL7DiJet, CMS7DiJet, and ATL13DiJet datasets are quite similar across different theoretical treatments, reflecting the behavior observed in inclusive jet measurements.
The only exception is the CMS8DiJet data, where the original NNLO theory (\textbf{Method 1}) shows a slight difference compared to the alternative treatments (\textbf{Method 2, 4, 5}). In contrast, \textbf{Methods 2, 4, and 5} yield similar impacts on the PDF error bands.
Compared to the inclusive jet case, the central values exhibit a greater discrepancy among various theoretical treatments in the dijet datasets. For example, the ATL13DiJet data impose a stronger constraint when using \textbf{Method 1}, as opposed to \textbf{Methods 2, 4}, and \textbf{5}, with the difference reaching approximately 3\% at $x \sim 0.4$.
However, the differences among \textbf{Methods 2, 4,} and \textbf{5} remain relatively small, around 1\%. 
When analyzing the ATL7DiJet, CMS7DiJet, and CMS8DiJet datasets, we observe a minor variation of approximately 1\% between the different theoretical treatments at $x \sim 0.4$, which is significantly smaller than the large PDF uncertainties associated with this $x$-region. 
Although the discrepancy may slightly increase at larger values of $x$ (e.g., $x > 0.4$), its impact on the PDFs can be disregarded due to the substantial PDF uncertainty in this extrapolated region. 
Applying different theoretical treatments to the dijet datasets results in only minor variations in the central values of the updated PDFs, with minimal influence on the PDF error bands. Specifically, \textbf{Methods 2, 4,} and \textbf{5} produce very similar gluon PDF ratios, with \textbf{Methods 4} and \textbf{5} being particularly close.
We conclude that applying different theoretical treatments to the inclusive jet datasets produces similar results for both the central values and the error bands. For dijets, a comparable pattern is observed in the error bands, although the central values display slightly greater variation, particularly in the extrapolated region where $x > 0.4$. For the remainder of this paper, we will adopt \textbf{Method 5} as our default approach, given its superior goodness of fit based on the $\chi^2/N_{\rm pt}$ criterion.

\subsection{Scale dependence }\label{Sec:ScaleImpact}

\begin{figure}[!h]
    \includegraphics[width=0.49\linewidth]{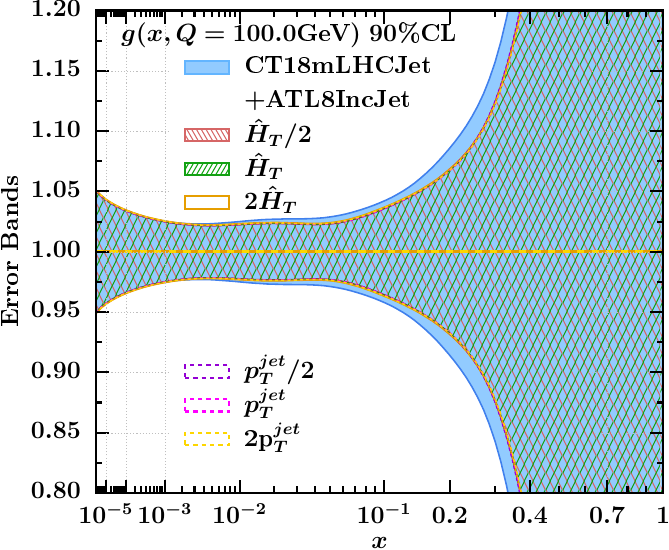}
    \includegraphics[width=0.49\linewidth]{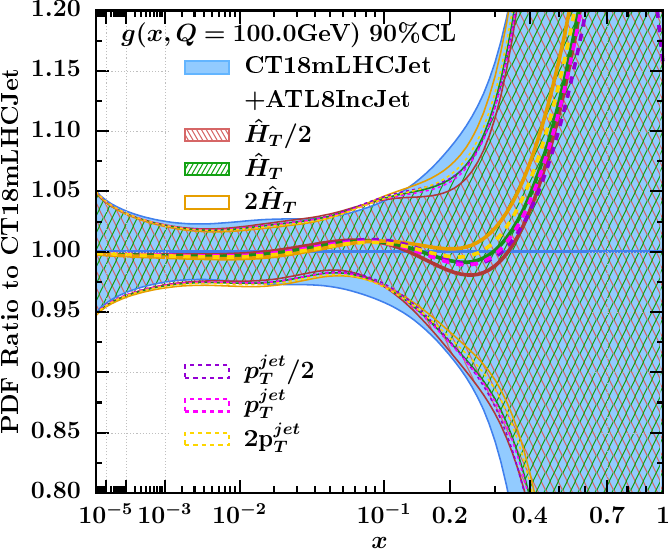}
    \includegraphics[width=0.49\linewidth]{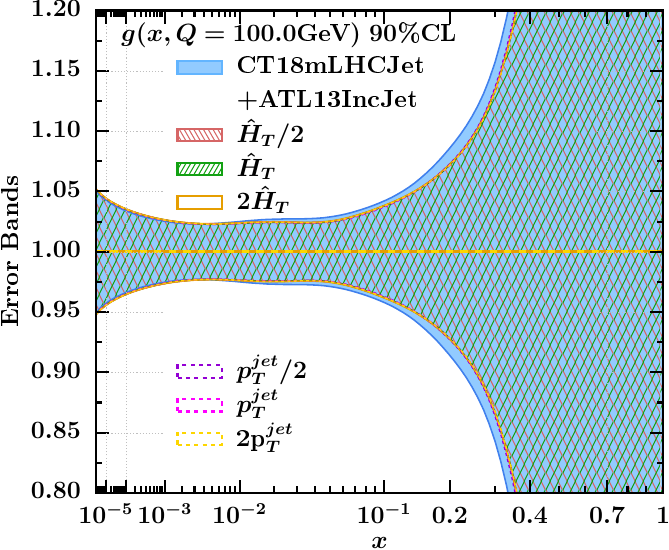}
    \includegraphics[width=0.49\linewidth]{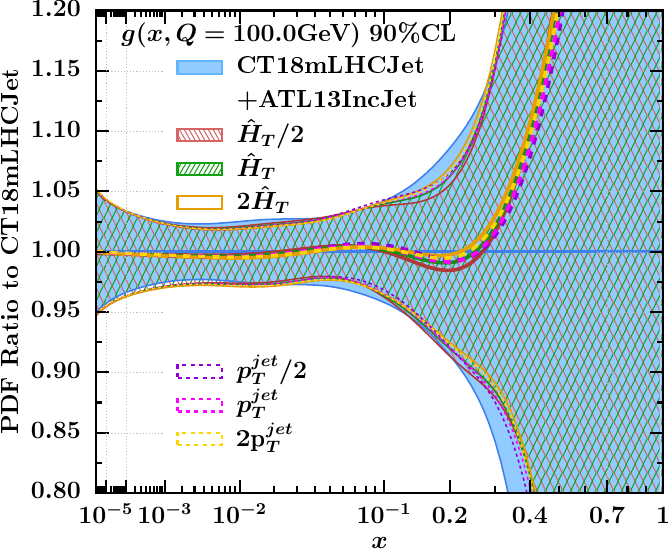}
    \includegraphics[width=0.49\linewidth]{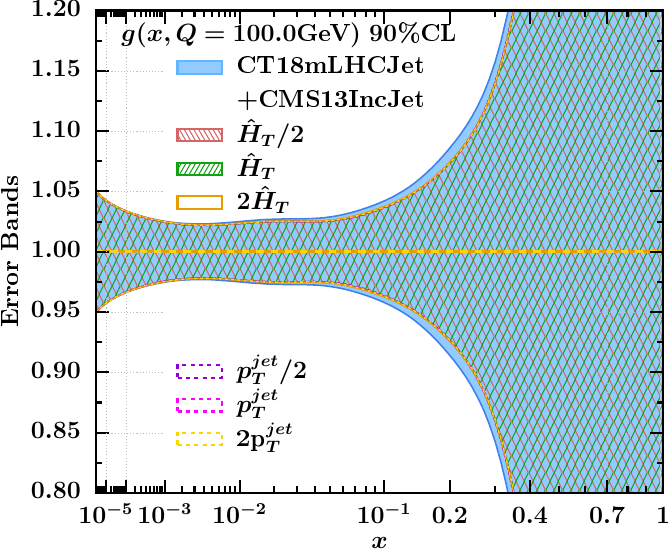}
    \includegraphics[width=0.49\linewidth]{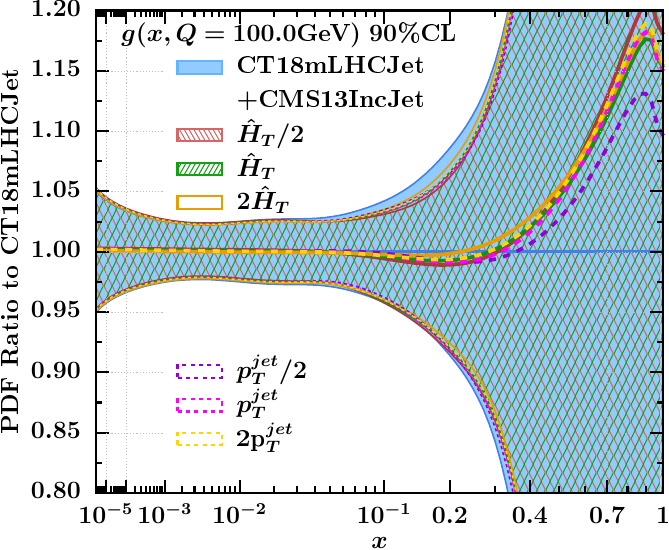}
\caption{The impact of each inclusive jet dataset on the gluon PDF at \HT and \ptj scales, along with scale variations, is analyzed using \texttt{ePump} after adding the ATL8IncJet, ATL13IncJet, and CMS13IncJet datasets to the CT18mLHCJet baseline one by one. The left plots show the relative gluon PDF uncertainty bands at a 90\% confidence level, while the right plots display the ratio of the new gluon PDF to the CT18mLHCJet.}
\label{Fig:ScaleVarInc}
\end{figure}

\begin{figure}[!h]
    \includegraphics[width=0.49\linewidth]{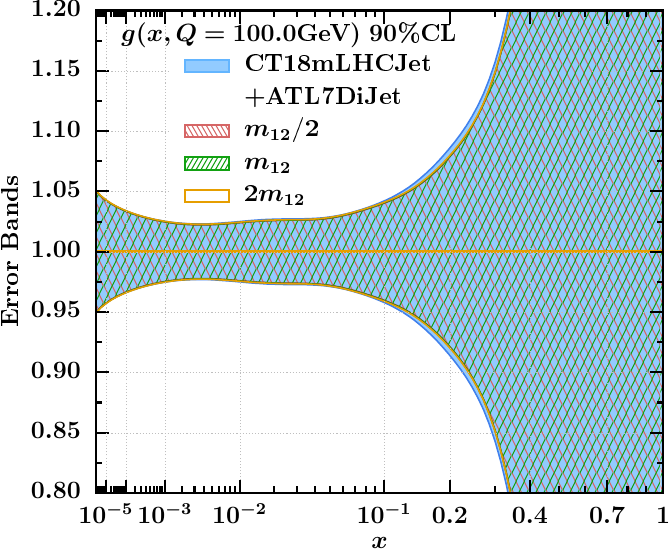}
    \includegraphics[width=0.49\linewidth]{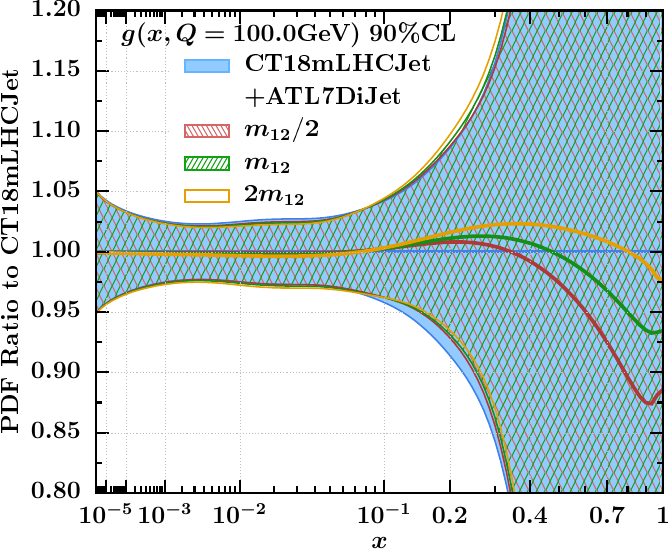}
    \includegraphics[width=0.49\linewidth]{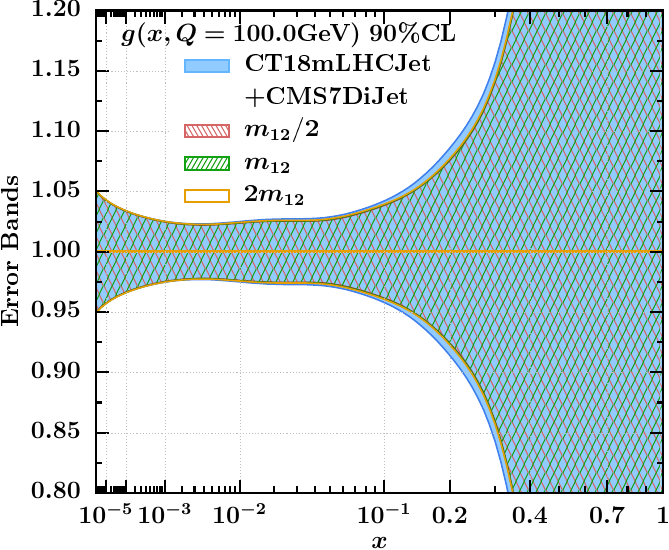}
    \includegraphics[width=0.49\linewidth]{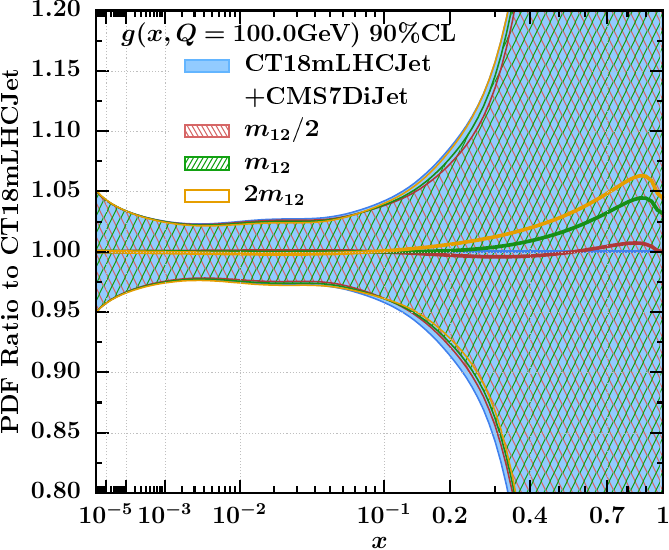}
    \includegraphics[width=0.49\linewidth]{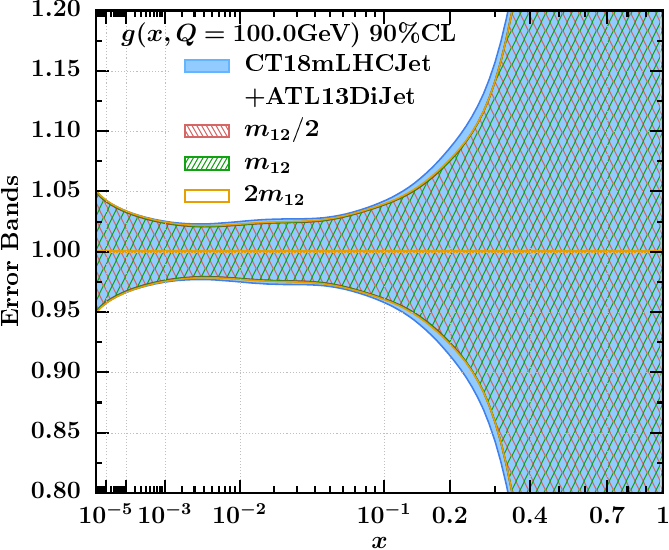}
    \includegraphics[width=0.49\linewidth]{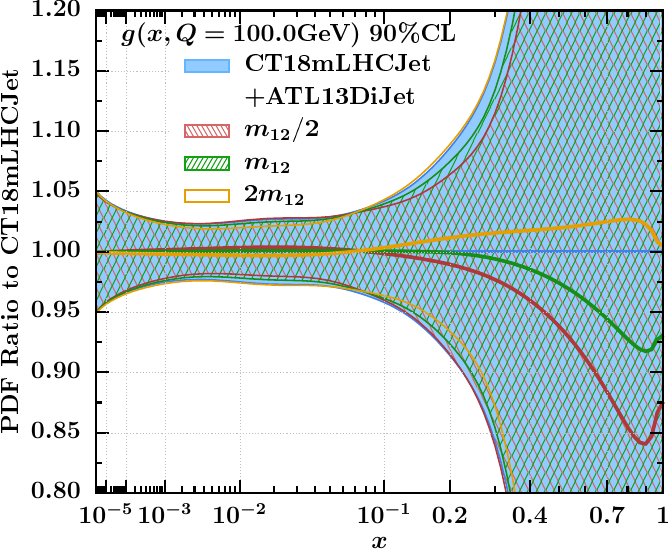}
    \caption{Same as Fig. \ref{Fig:ScaleVarInc}, but for dijet (ATL7DiJet, CMS7DiJet and ATL13DiJet datasets) with \mjj scale.}
\label{Fig:ScaleVarDiJet}
\end{figure}

\begin{figure}[!h]
    \includegraphics[width=0.49\linewidth]{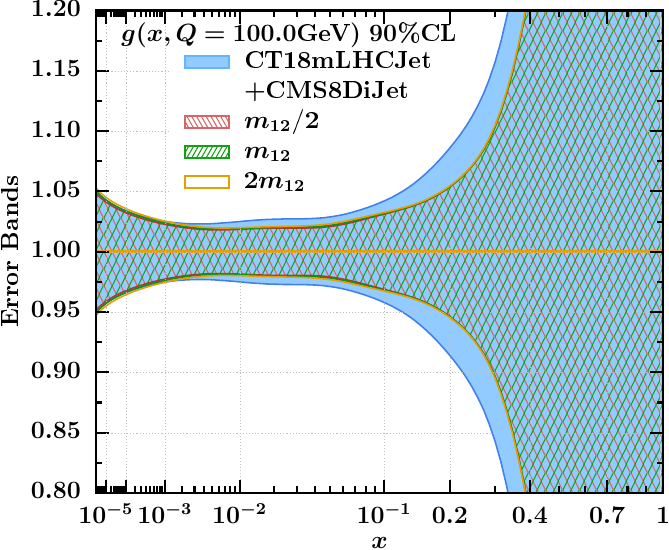}
    \includegraphics[width=0.49\linewidth]{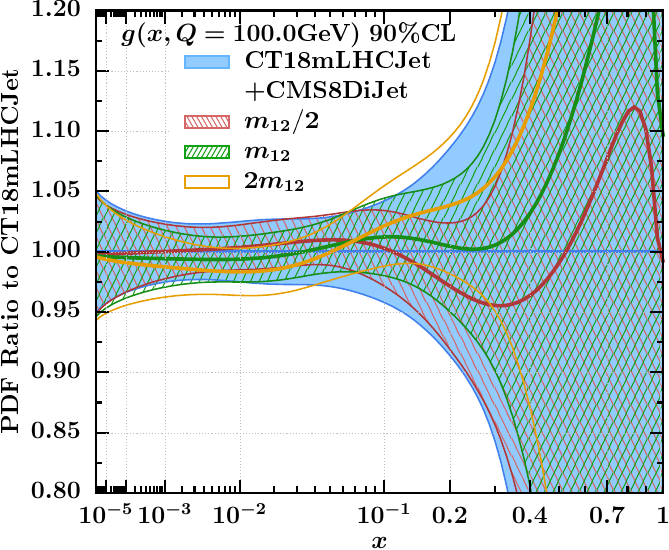}
    \includegraphics[width=0.49\linewidth]{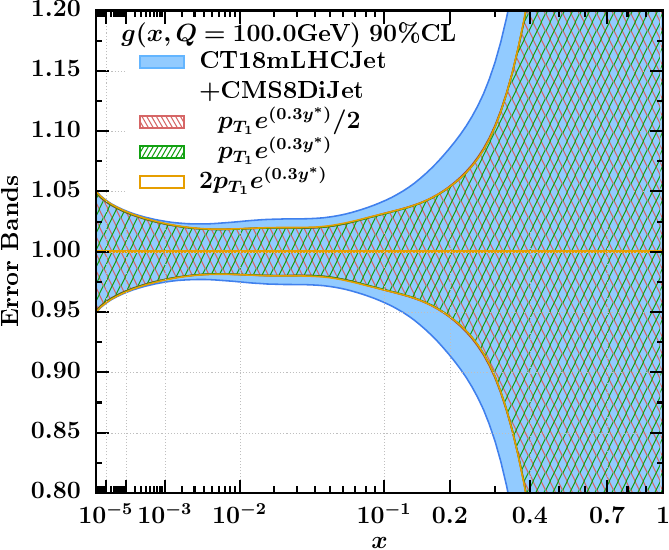}
    \includegraphics[width=0.49\linewidth]{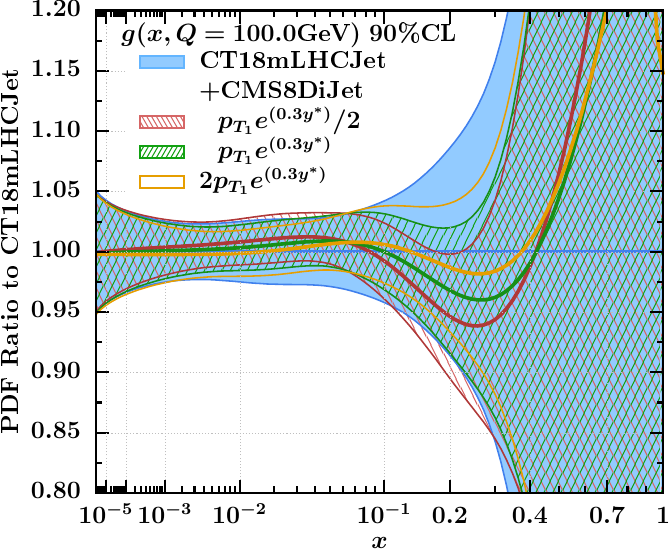}
\caption{Same as Fig. \ref{Fig:ScaleVarCMS8DiJet}, but for CMS8DiJet data with \mjj and \ptmax scales. }
\label{Fig:ScaleVarCMS8DiJet}
\end{figure}

In this section, we assess the impact of the central renormalization and factorization scales (\HT and \ptj for inclusive jets, and \mjj or \ptmax for dijets), along with the associated scale variations, on the CT18mLHCJet PDFs using the \texttt{ePump} method~\cite{Schmidt:2018hvu,Hou:2019gfw}.
The theoretical predictions for inclusive jet production at 7, 8, and 13 TeV are obtained with the convolution of the
 \texttt{APPLfast} tables~\cite{Britzger:2022lbf} with the chosen PDF sets, allowing for independent variation of the scales $\mu_R$ and $\mu_F$.
We can assess the impact of each dataset on the baseline PDF set, CT18mLHCJet, by adding them one by one using \texttt{ePump}.
The corresponding $\chi^2/N_{\rm pt}$ values after \texttt{ePump} updating for both the inclusive jet and dijet datasets are shown in Table
\ref{tab:Chi2ePumpIncJet} and Table \ref{tab:Chi2ePumpDijet}, 
respectively. The updated gluon PDFs at $Q=100~\GeV$ are presented in Fig. \ref{Fig:ScaleVarInc} for the inclusive jet and in Fig. \ref{Fig:ScaleVarDiJet} for the dijet.

From the impact of the inclusive jet datasets (ATL8IncJet, ATL13IncJet, CMS13IncJet) on the gluon PDF, as shown in Fig. \ref{Fig:ScaleVarInc}, we observe that both the \HT and \ptj scales, along with their variations, produce similar error bands.
We also observe that for all the inclusive jet datasets, the use of the \ptj scale and its variations produces similar central PDFs, with only a minor difference noted in the extrapolation region for the CMS13IncJet data.
The \HT scale, however, exhibits a slightly larger variation, most notably around $x \sim 0.3$, where the difference reaches approximately 1\% for ATL13IncJet and CMS13IncJet, and about 1.5\% for ATL8IncJet. These differences are negligible in all other regions.

Similarly, we show the impact of dijet datasets (ATL7DiJet, CMS7DiJet, ATL13DiJet and CMS8DiJet) on gluon PDFs in Figs. \ref{Fig:ScaleVarDiJet} and \ref{Fig:ScaleVarCMS8DiJet}, which indicate that both $m_{12}$ and \ptmax scales and variations yield similar error bands. However, the impact on the central PDF differs significantly across scale variations for dijet datasets. For the ATL13DiJet data set, the \mjjxt scale gives a harder gluon distribution,  while the \mjj and \mjjot scales prefer a softer gluon. 
The most pronounced difference is observed in CMS8DiJet data set around $x \sim 0.3$, where the central gluon PDF is unchanged for  \mjj scale, while it becomes softer and harder for \mjjot and $2m_{12}$, with the size reaching approximately 4\%, much larger than the one in the inclusive jet case.

Ideally, renormalization and factorization scales, as auxiliary quantities, should not influence theoretical predictions at all orders in perturbation theory. In a PDF global analysis, datasets with smaller scale dependence are preferred, as they reduce the bias introduced by the choice of scale in theoretical calculations. Comparatively, inclusive jet production shows relatively low sensitivity to renormalization and factorization scales, whereas dijet production, particularly the CMS8DiJet data, exhibits a strong scale dependence. This may stem from the more complex kinematics involved in dijet events, which can amplify the effects of scale choices on theoretical predictions. While this dataset has the potential to significantly reduce PDF error bands, further investigation into its scale dependence is necessary before it can be confidently included in final PDF fits.

Regarding the $\chi^2/N_{\rm pt}$ values in Table \ref{tab:Chi2ePumpIncJet}, the \ptj scale yields slightly better-fit quality than the \HT scale after applying the smooth KF and Monte Carlo error corrections 
(\textbf{Method 5}). However, in terms of pure NNLO predictions (\textbf{Method 1}), the \HT scale produces better $\chi^2/N_{\rm pt}$ values. The scale variations indicate that different theoretical treatments can lead to different optimal scale choices.
Taking the ATL8IncJet dataset as an example, \textbf{Method 1} prefers the \HTxt scale, while \textbf{Method 5} indicates that \HTot yields a better $\chi^2/N_{\rm pt}$ value. Similarly, for the \ptj scale variation, \textbf{Method 1} favors \ptjxt, whereas \textbf{Method 5} identifies \ptj as the optimal choice.
This instability poses a significant challenge in selecting an optimal scale. Fortunately, for inclusive jet datasets, regardless of the different treatments of a theory or scale variations, the impacts on the best-fit PDF central values and error bands remain largely consistent.

We conclude that inclusive jet data exhibit minimal sensitivity to the choice and variation of the central scale, making them ideal for constraining PDFs. In contrast, the dijet datasets, particularly CMS8DiJet, pose a significant challenge for global analyses due to their strong scale dependence. This highlights the need for additional comprehensive theoretical efforts before these datasets can be confidently included in final PDF fits, which will be addressed in future work. In the case of inclusive jets, although the optimal scale choice depends on the method of theoretical treatment, its impact on the resulting PDFs remains remarkably consistent.

\subsection{ Robustness of PDF constraints }\label{Sec:Opt}

\begin{figure}[h]
    \includegraphics[width=0.55\textwidth]{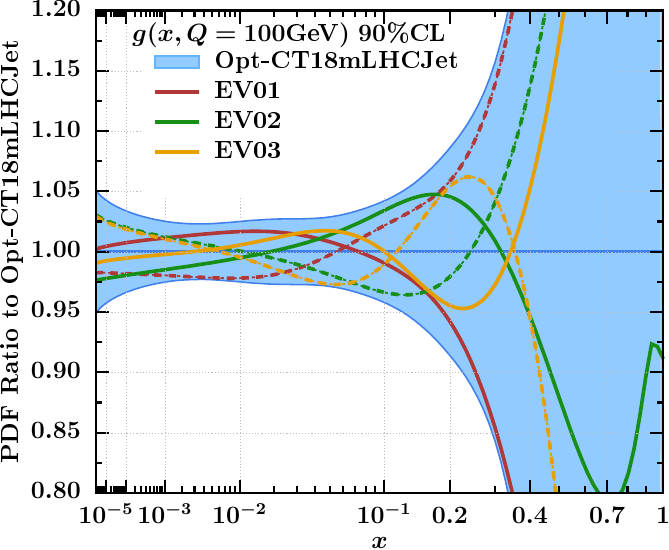}
    \caption{ The first three error PDF pairs corresponding to   eigenvectors EV01-EV03, obtained from applying the \texttt{ePump}-optimization procedure for the CT18mLHCJet+IncJetDiJet analysis. }
    \label{Fig:Opt}
\end{figure}

Due to the lack of information on the statistical correlation between the inclusive jet datasets~\cite{ATLAS:2017kux,ATLAS:2017ble,CMS:2021yzl} and the dijet datasets~\cite{ATLAS:2013jmu,CMS:2012ftr,CMS:2017jfq,ATLAS:2017ble}, both derived from the same collision events at the LHC, it is necessary to select one of these two types of data from each experiment to avoid double-counting the impact from the same source.
Below, we compare the impact of inclusive jet and dijet datasets on constraining the baseline CT18mLHCJet PDFs. For simplicity, we focus on datasets collected by the same experimental collaboration at the same collider energy and integrated luminosity. This can be achieved by performing the \texttt{ePump}-optimization procedure~\cite{Schmidt:2018hvu,Hou:2019gfw} on the CT18mLHCJet, incorporating the ATLAS and CMS datasets collected at 7 and 13 TeV. This results in new PDF error sets, referred to as the Opt-CT18mLHCJet PDFs. This optimization allows us to identify a reduced set of error PDFs that captures the majority of the PDF dependence of the observables under consideration.
The new eigenvectors retain the same information as the original ones but are optimized such that a smaller set of error PDFs can capture the required PDF sensitivity for a given set of observables. This allows us to evaluate and validate which datasets impose the strongest constraints on PDF uncertainties. Figure \ref{Fig:Opt} 
illustrates the behavior of the optimized eigenvector directions in the Opt-CT18mLHCJet analysis, while the fractional contributions to the PDF errors from the leading eigenvectors for each dataset are presented in Table \ref{tab:Opt}.

The optimization results are shown in Table \ref{tab:Opt}, the first four eigenvectors\footnote{Each eigenvector corresponds to two PDF error sets, representing positive and negative directions.}, out of a total of eight, account for approximately 98\% of the gluon-PDF error band. Notably, the contributions from the ATLAS 7 TeV, CMS 7 TeV, and ATLAS 13 TeV inclusive jet datasets are approximately 19.75\%, 17.64\%, and 21.84\%, respectively. In contrast, the dijet datasets contribute only 6.87\%, 11.78\%, and 17.63\%, respectively. This indicates that inclusive jet datasets generally impose stronger constraints on the gluon PDFs compared to dijet datasets at the same energy.

Additionally, as illustrated in Fig. \ref{Fig:Opt}, the optimized eigenvector sets play complementary roles in constraining the PDF error bands across different $x$ ranges. The first eigenvector (EV01, corresponding to Sets 1 and 2 in the optimized error sets) primarily constrains the regions of $x \in [10^{-3}, 10^{-2}]$ and $x > 0.3$, while the second eigenvector (EV02, corresponding to Sets 3 and 4) predominantly constrains the region around $x \sim 0.1$.

The \texttt{ePump} optimization presented in this section indicates that the inclusive jet datasets play a significant role in updating the CT18 PDFs, despite yielding a higher $\chi^2/N_{\rm pt}$ compared to the dijet datasets. The key points can be summarized as follows:

\begin{itemize}
\item \textbf{Theory treatment:}
As discussed in Sec.~\ref{Sec:TheoryTreatment}, the Monte Carlo error in the NNLO theory calculations is found to be non-negligible and must be accounted for in the PDF fits. Our analysis demonstrates that inclusive jet data are less sensitive to variations in the treatment of theory calculations.
\item \textbf{Scale dependence:}  
The analysis in Sec. \ref{Sec:ScaleImpact} reveals that inclusive jet datasets are less sensitive to the choice of scale compared to dijet datasets. This characteristic is particularly desirable, as it reduces the bias introduced by arbitrary scale choices in theoretical calculations.
\item \textbf{Robustness of PDF constraints:} 
When addressing potential double-counting from the inclusion of different datasets based on the same experimental collision events, it is crucial to select datasets that provide strong constraints on the PDFs. Our optimization study in this subsection suggests that inclusive jet datasets impose stronger constraints on the gluon PDF error band compared to dijet datasets, in agreement with findings from MSHT~\cite{Cridge:2023ozx} and NNPDF~\cite{AbdulKhalek:2020jut}.

\end{itemize}

\begin{table}[h]
\caption{Fractional contributions to the gluon PDF uncertainty from the leading eigenvectors (EVs), obtained through the \texttt{ePump} optimization procedure described in the text, are provided for the individual ATLAS and CMS inclusive jet and dijet datasets collected at 7 TeV and 13 TeV at the LHC. The first column presents the EV number, while the second column gives the percentage contribution of each EV to the gluon PDF error at a given $x$ value for $Q=100$ GeV. Contributions from the individual datasets are listed in the subsequent columns.}
\resizebox{\textwidth}{!}{
    \begin{tabular}{c|c|cc|cc|cc}
    \hline
     EV  No.    &  Percentage (\%)       &   CMS7IncJet & CMS7DiJet     &     ATL7IncJet    &     ATL7DiJet &   ATL13IncJet  &   ATL13DiJet  \\
    \hline
    1   & 65.04   &   12.88           & 4.40                 &     12.32              &     9.28               &   13.68             &   12.47            \\
    2   & 20.51   &   4.30            & 1.56                 &     3.55               &     1.88               &   5.24              &   3.98             \\
    3   & 9.97    &   2.57            & 0.91                 &     1.77               &     0.62               &   2.92              &   1.18             \\
    4   & 2.77    &   0.72            & 0.22                 &     0.52               &     0.11               &   0.90              &   0.30              \\
    \hline
    \end{tabular}
}
\label{tab:Opt}
\end{table}

\section{The Global analysis}\label{Sec:Globalfit}

We are now ready to include the desired inclusive jet data in the global analysis within the CT18 framework. We will present the impact of each new dataset on the gluon PDF by incorporating them one by one, and finally, all together, on top of the CT18 baseline (instead of CT18mLHCJet). Subsequently, we will evaluate the impact on gluon-gluon ($gg$) parton luminosities and their uncertainties, as well as the implications for the production of top quark pairs, inclusive Higgs, and associated $t\bar{t}H$ production at the LHC 14 TeV.

\subsection{The impact of new inclusive-jet data on the $g$ PDF}
\label{Sec:FitIncJet}
\begin{figure}[!h]

    \includegraphics[width=0.49\linewidth]{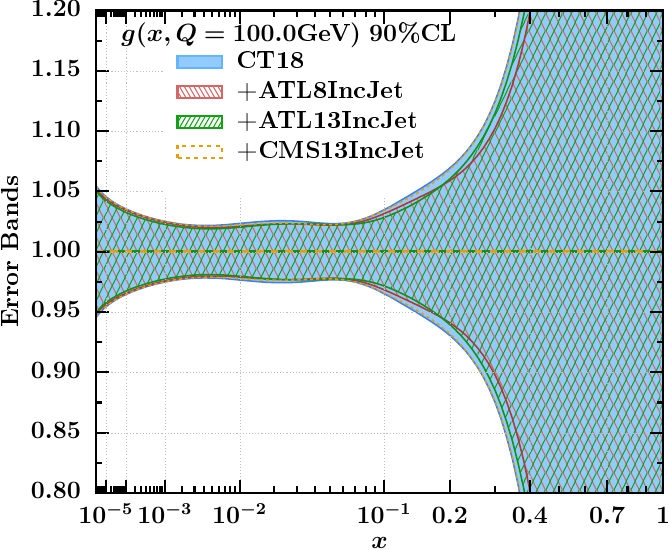}
    \includegraphics[width=0.49\linewidth]{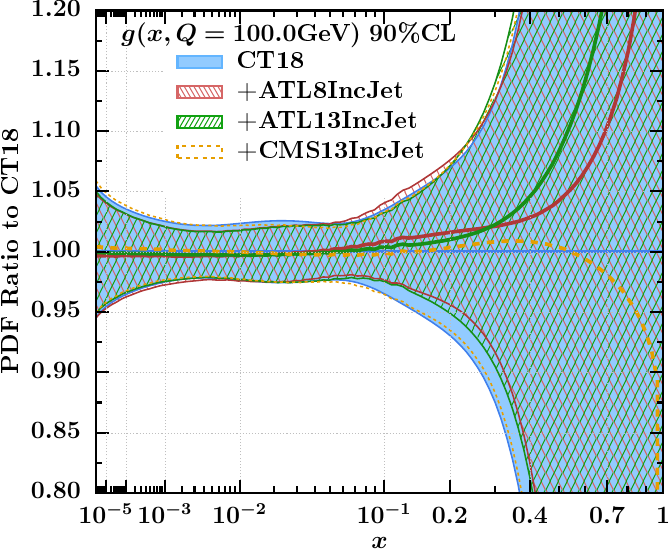}
    \includegraphics[width=0.49\linewidth]{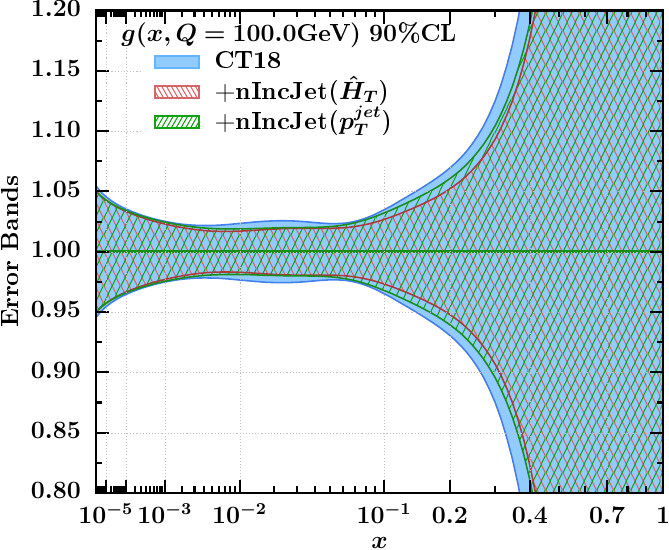}
    \includegraphics[width=0.49\linewidth]{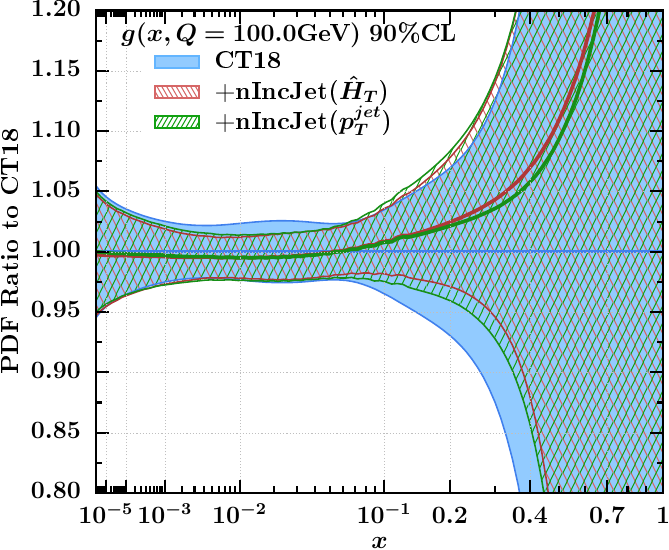}
    \caption{ The upper plots illustrate the impact of adding the inclusive jet datasets (ATL8IncJet, ATL13IncJet, and CMS13IncJet) to the CT18 baseline individually. The lower plots show the combined impact of all new inclusive jet datasets on the CT18 gluon PDF. The left panels depict the error bands at 90\% C.L., while the right panels show the ratios relative to the CT18 central value.}
    \label{Fig:GlobalFitCT18IncJet}
\end{figure}

In Fig. \ref{Fig:GlobalFitCT18IncJet}, we present the global fit results by adding each inclusive jet dataset \cite{ATLAS:2017kux,ATLAS:2017ble,CMS:2021yzl} on top of the CT18 baseline in the upper panel plots, and the combined impact of all new inclusive jet datasets in the lower panel plots. The corresponding $\chi^2/N_{\rm pt}$ are listed in Tab.~\ref{tab:Chi2GlobalFit-IncJet}. As previously discussed, since the scale choices \HT and \ptj yield very similar results, we focus on the \ptj scale choice in the upper panel and compare the differences in the lower panel plots.
We observe that the ATLAS 8 TeV and 13 TeV inclusive jet datasets prefer a harder gluon distribution compared to CT18 for $x > 0.1$. Similarly, the CMS 13 TeV inclusive jet data also pulls the gluon PDF toward a harder distribution at $x \sim 0.4$, although the effect is somewhat weaker compared to the ATLAS datasets. In general, the gluon PDF uncertainty is large at high $x$ due to the lack of direct experimental constraints in this region.

\begin{table}[!h]
	\caption{The global fit $\chi^2/N_{\rm pt}$ values for the inclusive jet datasets, added sequentially and simultaneously on top of the CT18 baseline, using both \HT and \ptj scale choices.}
	\begin{tabular}{lc|C{2cm}C{2cm}|C{2cm}C{2cm}}
 \hline
		\multirow{2}{*}{Data} &      \multirow{2}{*}{~~$N_{\rm pt}$~~}   & \multicolumn{2}{c|}{  one-by-one }      &   \multicolumn{2}{c}{ Optimal combination  }        \\
		\cline{3-6}
		 &                                       & \ptj         & \HT                 & \ptj               & \HT        \\
		\hline
		ATL8IncJet              &          171                          & 
		1.53               &   1.59            & 1.53                 &     1.58                                              \\
		ATL13IncJet             &          177                          & 1.24               &   1.29            & 1.24                 &     1.29                                               \\
		CMS13IncJet             &          78                           & 1.09               &   1.15            & 1.09                 &     1.16                                               \\

  \hline
	\end{tabular}
	\label{tab:Chi2GlobalFit-IncJet}
\end{table}

\begin{figure}[!h]
    \includegraphics[width=0.99\linewidth]{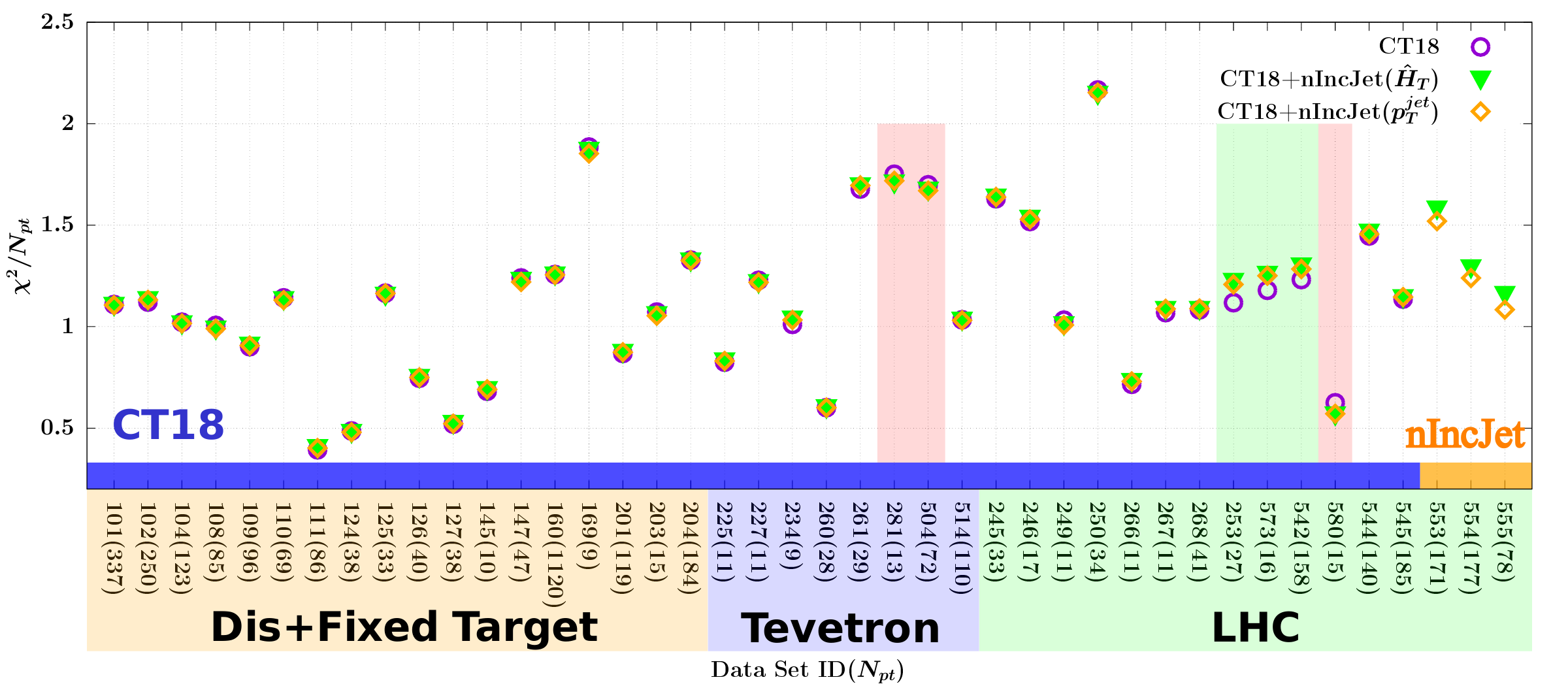}
    \caption{Comparison of $\chi^2/N_{\rm pt}$ values for individual datasets in CT18 and CT18+nJet fits after including inclusive jet datasets simultaneously for \HT and \ptj scales. The horizontal axis represents the dataset ID (number of data points), while the vertical axis displays the $\chi^2/N_{\rm pt}$ values. This plot illustrates the changes in $\chi^2/N_{\rm pt}$ for various experiments after the simultaneous inclusion of the new inclusive jet datasets.}
    \label{Fig:Chi2Npt}
\end{figure}

\begin{figure}[!h]

    \includegraphics[width=\linewidth]{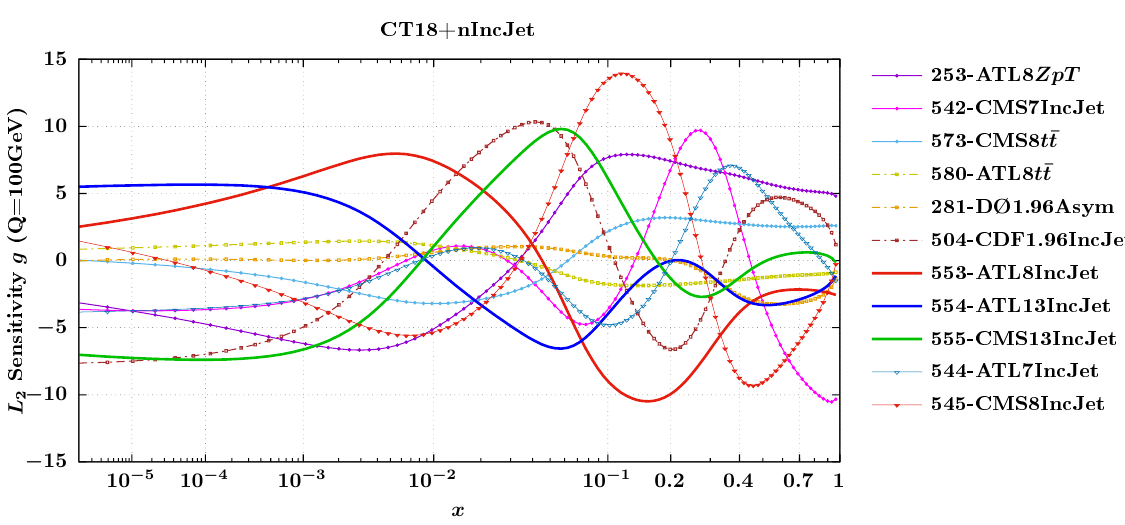}
    \caption{$L_2$ sensitivity plot for the gluon PDF at $Q=100$~GeV.
}
    \label{Fig:L2Sensitivity}
\end{figure}

\begin{table}[htbp]
	\caption{ 
 Comparison of the $\chi^2/N_{\rm pt}$ values for inclusive jet (dijet) datasets included in the CT18+nIncJet (CT18mLHCJet+DiJet) analysis with those reported by the  NNPDF~\cite{AbdulKhalek:2020jut} and MSHT~\cite{Cridge:2023ozx} groups.
}
	\begin{tabular}{lccccccc}
		\hline\hline
		Expt.  & Ref.                  & $\sqrt{s}$[TeV] & $\mathcal{L}_{\rm int}[\textrm{fb}^{-1}]$ & $N_{\rm pt}$ & CT & MSHT & NNPDF  \\
		\hline
		\multicolumn{8}{c}{Inclusive Jet}                                                                                      \\
		\hline
		ATLAS & \cite{ATLAS:2014riz} & 7               & 4.5                & 140      & 1.46    & 1.55& 1.59\\
		CMS   & \cite{CMS:2014nvq}   & 7               & 5.0                & 158      & 1.28    & 1.02& 1.09\\
		ATLAS & \cite{ATLAS:2017kux} & 8               & 20.3               & 171      & 1.53    & 1.94& 3.22\\
		CMS   & \cite{CMS:2016lna}   & 8               & 19.7               & 185      & 1.14    & 1.83& 1.19               \\
		ATLAS & \cite{ATLAS:2017ble} & 13              & 3.2                & 177      & 1.24    & --         & --                  \\
		CMS   & \cite{CMS:2021yzl}   & 13              & 36.5               & 78       & 1.09    & --         & --                  \\
		\hline
 \multicolumn{8}{c}{DiJet}    \\
		        \hline
		        ATLAS & \cite{ATLAS:2013jmu} & 7               & 4.5                & 90        & 0.93     & 1.05& 1.95\\
		        CMS   & \cite{CMS:2012ftr}   & 7               & 5.0                & 54        & 1.65     & 1.44& 2.08\\
		        CMS   & \cite{CMS:2017jfq}   & 8               & 19.7               & 122       & 1.09     & 1.22& 2.21\\
		        ATLAS & \cite{ATLAS:2017ble} & 13              & 3.2                & 136       & 0.92     & --           & --                   \\
		        \hline
	\end{tabular}
	\label{tab:Chi2Full}
\end{table}

The corresponding $\chi^2/N_{\rm pt}$ values for both \HT and \ptj scales are compared in Table \ref{tab:Chi2GlobalFit-IncJet}  for the new inclusive jet data, and in Fig. \ref{Fig:Chi2Npt} for all the CT18 datasets.
The two scales yield similar $\chi^2/N_{\rm pt}$ values for all the CT18 datasets. However, for the new inclusive jet data, the \ptj scale provides a slightly better $\chi^2/N_{\rm pt}$.
The ATLAS8ZpT, (ID=253), CMS7IncJet (ID=542), and CMS8ttb (ID=573) exhibit a slight increase in chi2/Npt, suggesting a potential tension between these datasets and the new inclusive jet data.
The $\chi^2/N_{\rm pt}$ values of the ATLAS8ttbar (ID=580), DO1.96Asym (ID=281), and CDF1.96IncJet (ID=504) datasets have improved, indicating that these datasets exhibit a trend similar to the new inclusive jet datasets. This observation is reinforced by the results of the $L_2$ sensitivity analysis \cite{Wang:2018heo,Jing:2023isu}, as shown in Fig.~\ref{Fig:L2Sensitivity}. A similar figure is provided in Appendix~\ref{L2SenT210} for convenience, with $T^2=10$.
We observe that the negative $L_2$ sensitivity values in the $x \in [0.1,0.4]$ region for ATL8$t\bar{t}$ (ID=580), D\O{}1.96Asym (ID=281), and CDF1.96IncJet (ID=504) are consistent with a preference for a harder gluon. In contrast, the positive $L_2$ sensitivity values for ATL8ZpT (ID=253), CMS7IncJet (ID=542), and CMS8$t\bar{t}$ (ID=573) indicate a pull towards a softer gluon distribution.

Furthermore, we compare the $\chi^2/N_{\rm pt}$ values for inclusive jet and dijet datasets analyzed within the CTEQ-TEA (CT) (CT18+nIncJet/CT18mLHCJet+DiJet) and NNPDF~\cite{AbdulKhalek:2020jut}, as well as MSHT~\cite{Cridge:2023ozx}, global analyses in Table~\ref{tab:Chi2Full}. It is important to note that the $\chi^2/N_{\rm pt}$ values were obtained using the $p_T^{\text{jet}}$ scale in both this study and in MSHT~\cite{Cridge:2023ozx}. We observe that most dijet datasets give better-fit quality than the inclusive jet datasets, consistent with the findings of MSHT~\cite{Cridge:2023ozx}. However, in contrast, NNPDF shows better-fit quality for the inclusive jet datasets than for the dijet datasets.

\subsection{Phenomenological implications}
\label{sec:pheno}

We now examine the phenomenological implications of the inclusive jet datasets, focusing primarily on the gluon parton and related processes at the LHC, such as Higgs boson production, top-quark pair production, and the associated production of a Higgs boson with a top-quark pair($t\bar{t}H$).  

\begin{figure}[!h]
    \includegraphics[width=0.49\linewidth]{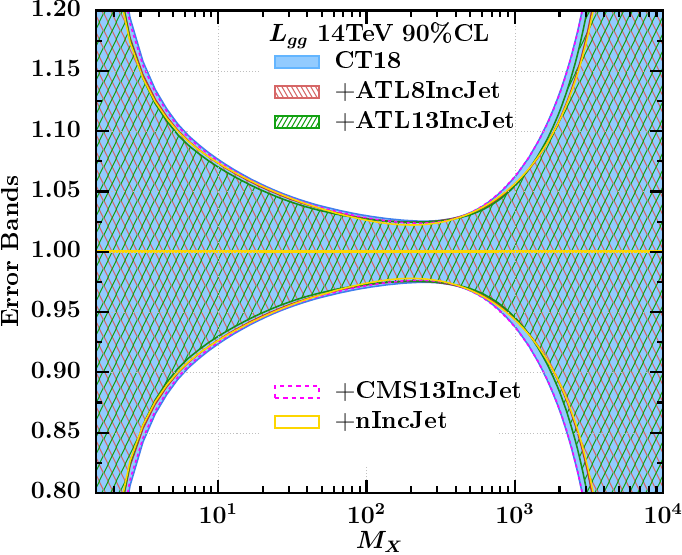}
    \includegraphics[width=0.49\linewidth]{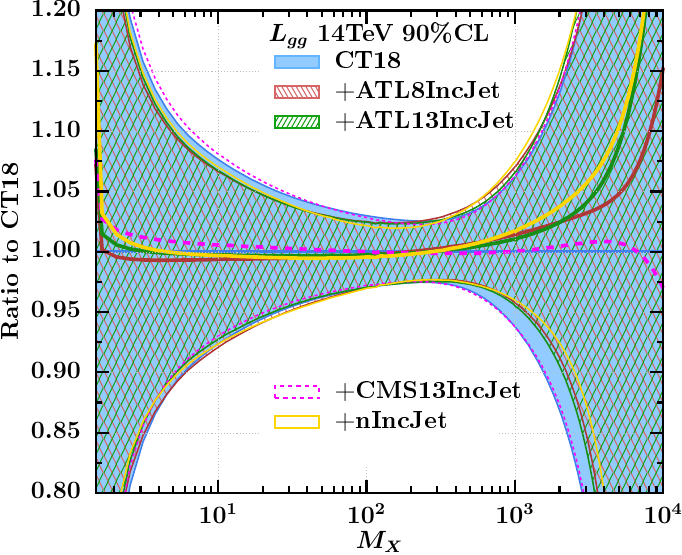}
    \caption{Comparison of the $gg$ parton luminosities (right) and their uncertainties (left) at the 14 TeV LHC, computed using the CT18 baseline and the CT18+ATL8IncJet, CT18+ATL13IncJet, CT18+CMS13IncJet, and CT18+nIncJet PDFs.
    }
    \label{Fig:PDFLuminosity}
\end{figure}

In Fig.~\ref{Fig:PDFLuminosity}, we compare the gluon-gluon ($gg$) parton luminosity $L_{gg}$ before and after incorporating the new inclusive jet datasets within the CT18 framework. 
Here the partonic luminosity is defined as~\cite{Campbell:2006wx} 
\begin{equation}\label{eq:Lgg}
L_{gg}(s,M_X^2)=\frac{1}{s}\int_{\tau=M_X^2/s}^1\frac{\dd x}{x}g(x,Q^2)g(\tau/x,Q^2),
\end{equation}
where the typical scale is chosen as the invariant mass $Q=M_X$. As shown, the PDF uncertainty on $L_{gg}$ is relatively reduced compared to CT18, highlighting the constraining power of the new inclusive jet datasets. In the region $M_X > 10^3$~GeV, the central value of $L_{gg}$ shifts in the harder direction, primarily due to the ATLAS 8 and 13 TeV inclusive jet datasets, which is consistent with the large gluon PDF shown in Fig.~\ref{Fig:GlobalFitCT18IncJet}.
In comparison, the impact of the CMS 13 TeV data is relatively weak, leaving 
$L_{gg}$ largely unchanged.
Overall, the final CT18+nIncJet fit accumulates the impact of the ATLAS 8 and 13 TeV datasets, resulting in the largest deviation from the CT18 reference.

\begin{figure}
    \centering
      \includegraphics[width=0.49\linewidth]{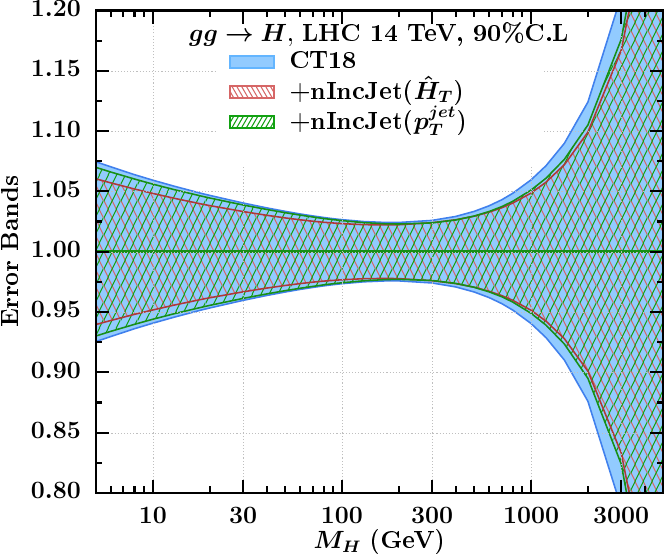}  
      \includegraphics[width=0.49\linewidth]{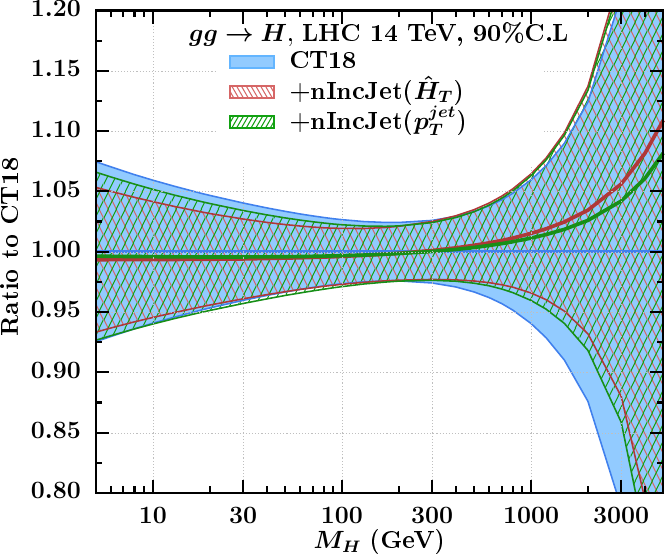}      
    \caption{Comparison of the CT18, CT18+nJet($\hat{H}_T$), and CT18+nJet$(p_T^{\rm jet})$ predictions (left) and the corresponding 90\% C.L. uncertainty (right) of the N3LO cross-section (right) for the Higgs-like scalar production in gluon fusion at the LHC 14 TeV, with varying the scalar mass $M_H$. 
}
    \label{Fig:N3LO-Calc-H}
\end{figure}

As an implication, we present the 14 TeV N3LO inclusive cross-sections for Higgs-like scalar production through gluon-gluon fusion in Fig.~\ref{Fig:N3LO-Calc-H},
while the Drell-Yan productions are collected in Fig.~\ref{Fig:N3LO-Calc-WZ} of Appendix~\ref{Sec:CT18mLHCJetvsCT18}.
The theoretical calculation is performed with the public code \texttt{n3loxs}~\cite{Baglio:2022wzu}, 
with the scalar particle mass $M_H$ varying from 5 GeV to 5 TeV, matching partonic invariant mass $M_X$ in Eq.~(\ref{eq:Lgg}).
For the mass at Higgs value $M_H=125~\GeV$, we have validated the calculation with \texttt{ggHiggs}~\cite{Bonvini:2016frm}, which shows a perfect agreement.
We see that the PDF uncertainty on the scalar production cross-section in Fig.~\ref{Fig:N3LO-Calc-H} is reduced in both the low and high mass regions, as expected.
In the high mass region, particularly when
$M_H>1$~TeV, the cross-section is noticeably enhanced, consistent with $L_{gg}$ in Fig.~\ref{Fig:PDFLuminosity}.
In comparison, the cross-section for small
$M_H$ values remain largely unchanged.
In Fig.~\ref{Fig:N3LO-Calc-H}, we also compare the impact from the fits with two jet scale choices, \HT and \ptj, respectively, which gives an overall consistent trend. That is to say, the central predictions are enhanced, and the PDF uncertainty is reduced. 
However, comparatively, the \HT
scale choice has a larger impact on both the uncertainty and the pull in the central value.
By comparison, the Drell-Yan cross-section in Fig.~\ref{Fig:N3LO-Calc-WZ}  exhibits similar behavior, albeit with a slightly smaller impact.

\begin{figure}[!h]
    \includegraphics[width=0.49\linewidth]{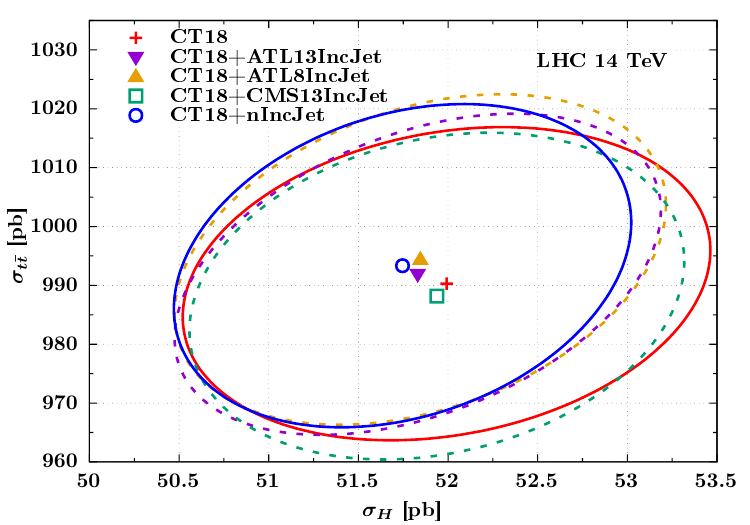}
    \includegraphics[width=0.49\linewidth]{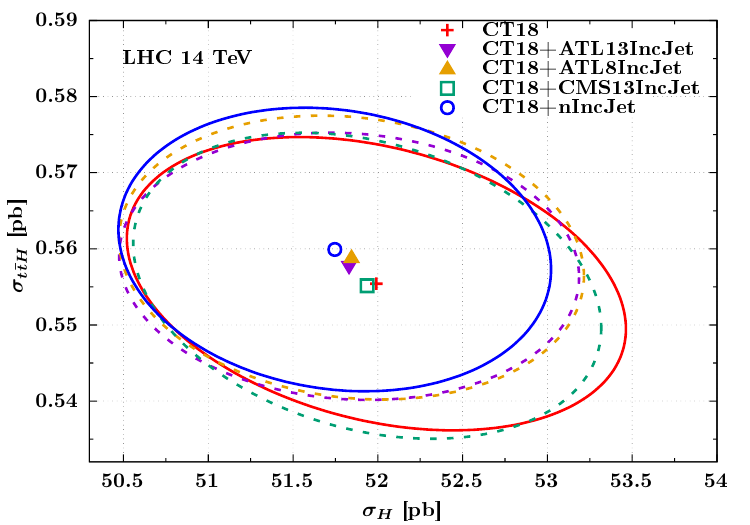}
    \caption{ Correlation ellipses for inclusive Higgs production in gluon fusion, $t\bar{t}$, and $t\bar{t}H$ production processes shown at 90\% C.L. 
    at the LHC  with $\sqrt s =  14$ TeV.}
     \label{Fig:CorrEllips}
\end{figure}

In a realistic scenario, we compare the correlation ellipses for Higgs production via gluon fusion, top-quark pair production, and associated Higgs-top-quark pair production at the LHC at 14 TeV in Fig.~\ref{Fig:CorrEllips}.
Similarly as before, the Higgs cross section is calculated at N3LO with \texttt{n3loxs}~\cite{Baglio:2022wzu} and Higgs mass is taken as $M_H=125~\GeV$.
The top-quark pair production is calculated with \texttt{Top++}~\cite{Czakon:2011xx} at NNLO, with soft gluons resummed up to NNLL, and the factorization and renormalization scales set to the top-quark mass, $m_t = 172.5$ GeV. The associated $t\bar{t}H$ production is calculated with \texttt{MadGraph\_aMC@NLO}~\cite{Alwall:2014hca,Frederix:2018nkq} at NLO, with the scale chosen as $M_T/2$, where $M_T$ sums of the transverse masses of the final-state particles.

\begin{figure}[!h]
    \includegraphics[width=0.32\linewidth]{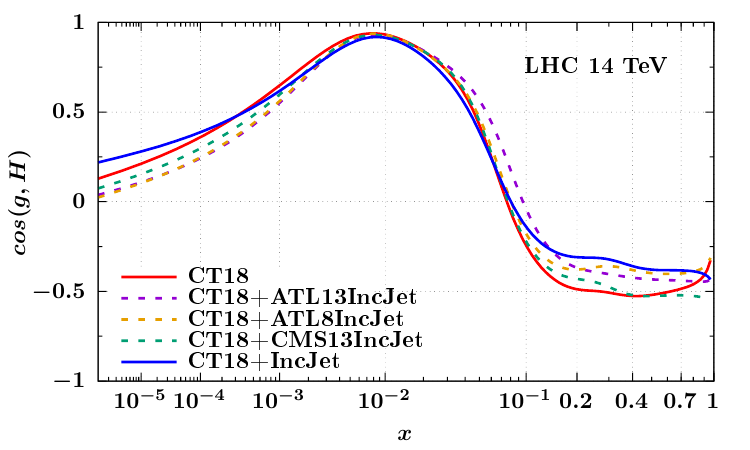}
    \includegraphics[width=0.32\linewidth]{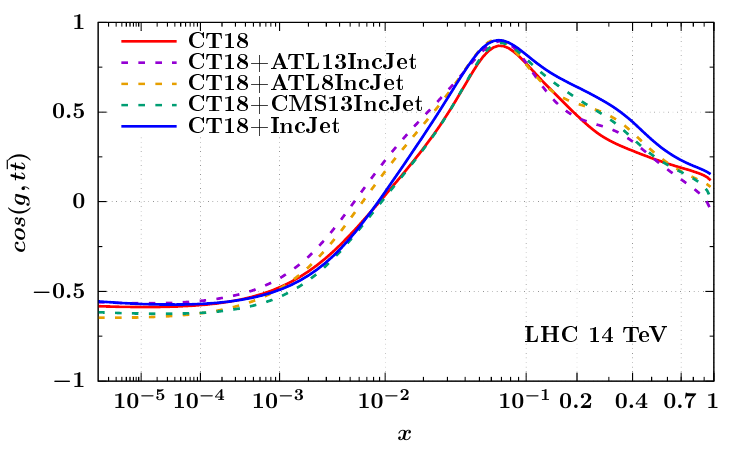}
    \includegraphics[width=0.32\linewidth]{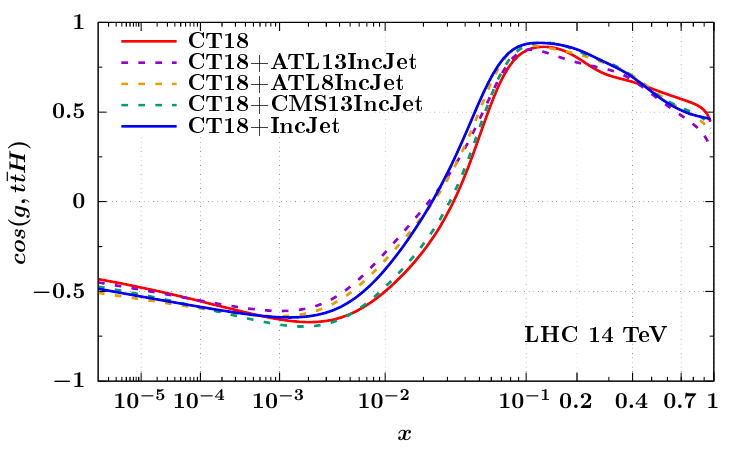}
    \caption{ Correlation cosine for inclusive Higgs production in gluon fusion, $t\bar{t}$, and $t\bar{t}H$ production processes shown at 90\% C.L. 
    at the LHC with a $\sqrt s=  14$ TeV.}
     \label{Fig:CorrCosine}
\end{figure}

As shown in Fig.~\ref{Fig:CorrEllips},
compared to the CT18 baseline, the CT18+nIncJet PDFs lead to a reduction in the Higgs production cross-section, whereas the cross-sections for the $t\bar{t}$ and $t\bar{t}H$ processes exhibit a slight increase.
This behavior can be explained by the correlation cosine \cite{Gao:2013bia} between the $H$, $t\bar{t}$ and $t\bar{t}H$  cross-sections, and the gluon PDF, as shown in Fig.~\ref{Fig:CorrCosine}.
We observe an anti-correlation between the Higgs production cross-section and the gluon PDF at $x \sim 0.3$, whereas the $t\bar{t}$ and $t\bar{t}H$ cross-sections positively correlate with the gluon PDF. The reduction of the error ellipses in Fig.~\ref{Fig:CorrEllips}  indicates a decrease in the uncertainties of the gluon PDF.

To conclude this section, the inclusion of the latest LHC inclusive jet measurements in the CT18+nIncJet PDF analysis results in a reduction of PDF uncertainties and an enhancement of the central value for the large-$x$ gluon PDF, as well as the $gg$ luminosity at large invariant masses. The phenomenological impact is assessed through theoretical predictions for gluon-initiated processes, leading to an increased cross section for the Higgs-like scalar at large mass, top-quark pair, and $t\bar{t}H$ production, while the cross section for inclusive Higgs boson production at the 14 TeV LHC shows a slight decrease.

\section{Conclusion and outlook}\label{Sec:Conclusion}

In this work, we have presented a comprehensive study of the impact of new LHC inclusive jet data~\cite{ATLAS:2017kux,ATLAS:2017ble,CMS:2021yzl}, as well as dijet data~\cite{ATLAS:2013jmu,CMS:2012ftr,CMS:2017jfq,ATLAS:2017ble}, on the CT18 PDFs, which is critical for the upcoming release of the next generation of CTEQ-TEA PDFs~\cite{Ablat:2024nhy}.
Before conducting time-intensive global fittings, we employed the fast Hessian profiling technique with \texttt{ePump}~\cite{Schmidt:2018hvu,Hou:2019gfw} to evaluate the impact of jet data.
This analysis is based on the profiled $\chi^2$ values and the resulting PDFs, including both their central values and uncertainty bands. To ensure the quality of the fit and reliable PDF constraints, we have thoroughly examined key experimental and theoretical aspects of the jet datasets.
\begin{itemize}[topsep=0pt]
\item \textbf{Decorrelation of systematic errors}. 
In Sec.~\ref{Sec:Overview}, we have addressed the systematic errors associated with ATLAS 8 and 13 TeV jets, which are not fully correlated across all rapidity jet bins. Following the ATLAS recommendation~\cite{ATLAS:2017kux}, we have explored the decorrelation of systematic uncertainties among different rapidity bins, into two or three subcomponents, with $\chi^2$ presented in Tab.~\ref{tab:DecorChi2ePump} and PDFs in Fig.~\ref{Fig:DecorImpactePump}.
We find that the two-subcomponent options have little impact on both $\chi^2$ and PDFs, while the three-subcomponent splittings lead to a significant improvement.
For the ATLAS 8 TeV inclusive jet data, the ATLAS recommendation, \emph{i.e.}, ``SuggestedR6", gives the best $\chi^2$, while ATLAS 13 TeV data prefer the decorrelation of Jet Energy Scale (JES) Pile-up Rho topology (Option 18).
In comparison, the impact on the PDF central values and error bands, as shown in Fig.~\ref{Fig:DecorImpactePump}, is minimal across all options, similar to the findings in ATLASpdf21~\cite{ATLAS:2021vod} and MSHT~\cite{Harland-Lang:2017ytb} fits.

\item \textbf{Monte Carlo statistical error}.
Motivated by previous studies conducted by the CTEQ-TEA~\cite{Hou:2019efy}, MSHT~\cite{Harland-Lang:2017ytb}, and NNPDF~\cite{Gehrmann-DeRidder:2016ycd,AbdulKhalek:2020jut} groups, we have examined in Sec.~\ref{Sec:TheoryTreatment} the various approaches to handling statistical errors in theoretical calculations. These include MC errors from integration, smoothing K-factors, and additional 0.5\% MC uncorrelated errors.
As shown in Tabs.~\ref{tab:Chi2ePumpIncJet}-\ref{tab:Chi2ePumpDijet}, we have achieved a reasonably good fit by simultaneously applying smooth K-factors and an additional 0.5\% uncorrelated MC uncertainty.
In terms of the resulted central PDFs and error bands in Fig.~\ref{Fig:DiffThImpactIncJet}, we have obtained a consistent impact of inclusive jet datasets~\cite{ATLAS:2017kux,ATLAS:2017ble,CMS:2021yzl}, independent of the treatments of the MC statistic errors. In contrast, the impact of dijet datasets~\cite{ATLAS:2013jmu,CMS:2012ftr,CMS:2017jfq,ATLAS:2017ble} in Fig.~\ref{Fig:DiffThImpactDijet} exhibits a strong dependence on the MC error treatments, especially the central value. 

\item \textbf{Scale Dependence}.
In Sec.~\ref{Sec:ScaleImpact}, we have analyzed the scale dependence of the theoretical predictions for the inclusive jet~\cite{ATLAS:2017kux,ATLAS:2017ble,CMS:2021yzl} and dijet~\cite{ATLAS:2013jmu,CMS:2012ftr,CMS:2017jfq,ATLAS:2017ble} datasets. 
By comparing Figs.~\ref{Fig:ScaleVarInc}-\ref{Fig:ScaleVarCMS8DiJet}, we see that the inclusive jet datasets show less sensitivity to scale choices and variations, underscoring their reliability for global PDF fits. In contrast, the dijet datasets, particularly the CMS 8 TeV dijet data~\cite{CMS:2017jfq} in Fig.~\ref{Fig:ScaleVarCMS8DiJet}, exhibit a strong scale dependence, warranting further theoretical investigation before being included in a global QCD analysis. This holds despite the triple-differential observable indicating a strong constraint on the gluon PDF in the large-\(x\) region. 
Additionally, if the ATLAS collaboration measures triple-differential dijet production in the near future, it could provide valuable validation of its constraining power for future PDF determinations.
\item \textbf{Robustness of PDF Constraints}.
Due to the lack of statistical correlation between the inclusive jet~\cite{ATLAS:2017kux,ATLAS:2017ble,CMS:2021yzl} and dijet~\cite{ATLAS:2013jmu,CMS:2012ftr,CMS:2017jfq,ATLAS:2017ble} datasets from the same collision events at the same LHC energy in the same detector, we need to select only one of them to avoid double counting. 
In Sec.~\ref{Sec:Opt}, we have included the inclusive jet and di-jet data sets simultaneously in the \texttt{ePump} optimization~\cite{Schmidt:2018hvu,Hou:2019gfw}  to examine the robustness.
In terms of the leading eigenvector sets in Fig..~\ref{Fig:Opt} and the fractional contribution presented in Tab.~\ref{tab:Opt}, we see that the inclusive jet datasets contribute more significantly to the optimization of gluon PDF uncertainties, indicating greater robustness compared to dijet datasets. 
This observation is consistent with findings in the NNPDF study~\cite{AbdulKhalek:2020jut}, where the inclusive jet leads to a more significant reduction of the gluon uncertainty.
\end{itemize}

Based on these observations, we have incorporated the new inclusive jet datasets into our optimal global fitting procedure, leading to the development of a new PDF set referred to as CT18+nIncJet, as discussed in Sec.~\ref{Sec:Globalfit}.
The $\chi^2/N_{\rm pt}$ values for the new inclusive jet datasets, evaluated using two typical scale choices, are compared in Tab.~\ref{tab:Chi2GlobalFit-IncJet}, while the resulting PDFs are shown in Fig.~\ref{Fig:GlobalFitCT18IncJet}.
Consistent with the \texttt{ePump} results in Sec.~\ref{Sec:ScaleImpact}, the \ptj scale yields a slightly better $\chi^2$ compared to $\hat{H}_T$, while the fitted PDFs remain largely independent of the scale choice in terms of both central values and uncertainty bands.
 We also have compared the $\chi^2$ values obtained with the \ptj scale to those from the NNPDF~\cite{AbdulKhalek:2020jut} and MSHT~\cite{Cridge:2023ozx} analyses in Tab.~\ref{tab:Chi2Full}, demonstrating overall consistency in the fitting quality.
The modest improvement in our global analysis is primarily driven by decorrelation and, in particular, the Monte Carlo error treatment described in Sec.~\ref{Sec:TheoryTreatment}, where we simultaneously include both smoothed K-factors and Monte Carlo integration errors in the theoretical calculations.
We have compared the gluon PDF, including both central values and uncertainty bands, from CT18 and CT18+nIncJet in Fig.~\ref{Fig:GlobalFitCT18IncJet}.
 The post-CT18 inclusive jet datasets favor a harder gluon PDF in the large-$x$ region, primarily driven by the ATLAS 8 and 13 TeV data. A slightly weaker pull from the CMS 13 TeV data suggests some tension, confirmed by the $L_2$ sensitivity analysis, as shown in Fig.~\ref{Fig:L2Sensitivity}. This tension is also observed in the MSHT~\cite{Cridge:2023ozx} and NNPDF~\cite{AbdulKhalek:2020jut} analyses. Overall, the reduction in the gluon PDF uncertainty bands shown in Fig.~\ref{Fig:GlobalFitCT18IncJet} highlights the constraining power of the new inclusive jet datasets.
Additionally, we have confirmed the dependence of the fitted PDFs, in terms of central values and uncertainty bands, on the scale choice is very mild, consistent with the \texttt{ePump} analysis in Sec.~\ref{Sec:ScaleImpact}.

To illustrate the implications of the Inclusive jet datasets, we provided gluon-gluon luminosity $L_{gg}$ In Sec.~\ref{sec:pheno} in Fig.~\ref{Fig:PDFLuminosity}. We observe an enhancement in the central value and a reduction in the error bands in the large invariant mass region. Furthermore, we have analyzed gluon-initiated processes at the LHC 14 TeV, including the inclusive Higgs(-like) scalar, top-quark pair, and $t\bar{t}H$ associated productions to demonstrate the LHC phenomenology. To amplify the impact of the gluon PDF, we first considered Higgs-like scalar production, with the cross-section as a function of the scalar particle mass up to 5 TeV, shown in Fig.~\ref{Fig:N3LO-Calc-H}. As a result, the overall features of the $L_{gg}$ luminosity in Fig.~\ref{Fig:PDFLuminosity} are fully inherited. 
As a realistic example, we present the correlation ellipse for the production cross-sections of a Higgs, top-quark pair, and $t\bar{t}H$  in Fig.~\ref{Fig:CorrEllips}.
The inclusive Higgs boson production cross-section exhibits a slight reduction, whereas the top-quark pair and $t\bar{t}H$ cross-sections increase mildly, highlighting the anti-correlation and correlation with the large-$x$ gluon PDF, as shown in Fig.~\ref{Fig:CorrCosine}.

 As a final remark, the overall impact of the inclusive jet and dijet datasets obtained in this study is consistent with the findings of the MSHT~\cite{Cridge:2023ozx} and NNPDF~\cite{AbdulKhalek:2020jut} studies.
 However, the optimal choice of inclusive jet datasets in this work deviates from the preferences suggested by MSHT~\cite{Cridge:2023ozx} and NNPDF~\cite{AbdulKhalek:2020jut}, primarily due to a slightly different underlying philosophy.
 In this work, along with the CTEQ-TEA series~\cite{Hou:2019efy,Sitiwaldi:2023jjp,Ablat:2023tiy}, we not only require a strong goodness-of-fit criterion~\cite{Kovarik:2019xvh} but also consistency in theoretical predictions. Consequently, we prefer the inclusive jet datasets over the dijet ones due to their weaker scale dependence, even though they exhibit a stronger impact on the gluon PDF.
In addition, the treatment of decorrelation and Monte Carlo errors investigated in this work improves the goodness-of-fit and reduces the tension among the inclusive jet datasets. Alongside other high-precision LHC measurements (\emph{e.g.}, the production of Drell-Yan~\cite{Sitiwaldi:2023jjp} and top-quark pair~\cite{Ablat:2023tiy}), this work is expected to provide a key input towards the upcoming release of a new generation of CTEQ-TEA PDFs~\cite{Ablat:2024nhy}.

\begin{acknowledgements}
We would like to thank our CTEQ-TEA colleagues for many helpful discussions. 
The work of S. Dulat and I. Sitiwaldi are supported by the National Natural Science Foundation of China under Grant No.11965020 and Grant No. 11847160, respectively.
The work of K. Xie and C.-P. Yuan was  supported by 
the U.S. National Science Foundation under Grants No.~PHY-2310291 and PHY-2310497.
P.M.N. was partially supported by the U.S. Department of Energy under Grant No.~DE-SC0010129. This work used the high-performance computing clusters from SMU M3 and the MSU HPCC.
\end{acknowledgements}

\appendix
\section{Supplementary material}\label{Sec:CT18mLHCJetvsCT18}

\subsection{Impact from individual jet data sets}

Here, we present comparisons between the CT18mLHCJet gluon PDF and the newly obtained gluon PDFs, each derived by incorporating one of the inclusive jet  datasets (ATL7Incjet, CMS7Incjet, ATL8Incjet, CMS8Incjet, ATL13Incjet, and CMS13Incjet) and dijet datasets (ATL7DiJet, CMS7DiJet, and CMS8DiJet) 
individually. The $\chi^2/N_{\rm pt}$ fitted individually and simultaneously for inclusive jet datasets are already presented in Tab.~\ref{tab:Chi2GlobalFit-IncJet} of Sec.~\ref{Sec:FitIncJet}. In this appendix, we list a similar setup for the dijet datasets with both \mjj and \ptmax scale choices on top of the CT18mLHCJet baseline in Tab.~\ref{tab:Chi2GlobalFit-dijet} for completeness.
The \mjj scale yields a slightly better $\chi^2/N_{\rm pt}$ except for CMS8DiJet. 
Overall, we obtain a reasonable goodness of fit, aligning with the MSHT result~\cite{Cridge:2023ozx}, but in contrast to the NNPDF one~\cite{AbdulKhalek:2020jut}.

\begin{table}[!h]
	\caption{The global fit $\chi^2/N_{\rm pt}$ values for the DiJet datasets, added sequentially and simultaneously on top of the CT18mLHCJet baseline, using both \mjj and \ptmax scale choices.}
	\begin{tabular}{lc|C{2cm}C{2cm}|C{2cm}C{2cm}}
 \hline
		\multirow{2}{*}{Data} &      \multirow{2}{*}{~~$N_{\rm pt}$~~}   & \multicolumn{2}{c|}{  one-by-one }      &   \multicolumn{2}{c}{ Optimal combination  }        \\
		\cline{3-6}
		                        &                                       & \mjj        & \ptmax                 & \mjj               & \ptmax        \\
		\hline
		ATL7DiJet               &          90                           & 0.80               &   -            & 0.81                 &     0.93                                              \\
		CMS7DiJet               &          54                           & 1.59               &   -            & 1.60                 &     1.65                                               \\
		CMS8DiJet               &          122                          & 1.19               &   1.03         & 1.20                 &     1.09                                               \\
        ATL13DiJet              &          136                          & 0.91               &   -            & 0.92                 &     0.93                                              \\

  \hline
	\end{tabular}
	\label{tab:Chi2GlobalFit-dijet}
\end{table}

\begin{figure}[htbp]

	\includegraphics[width=0.49\linewidth]{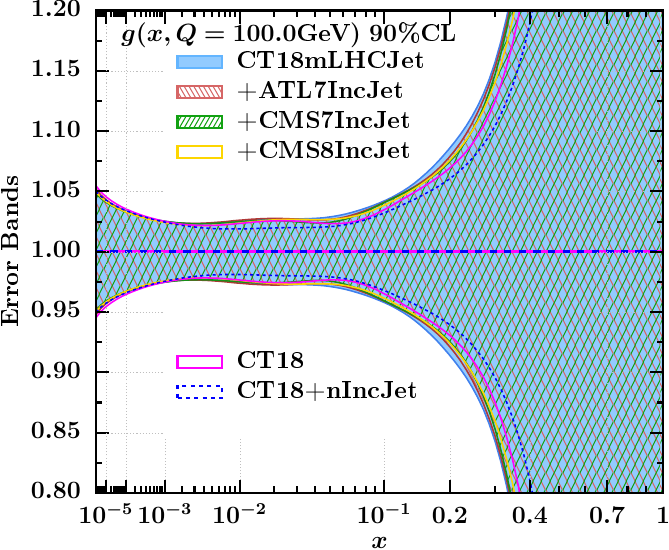}
	\includegraphics[width=0.49\linewidth]{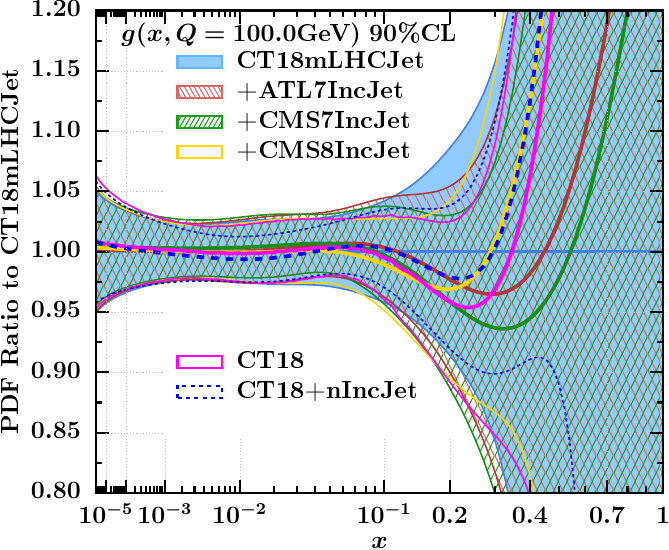}
	\includegraphics[width=0.49\linewidth]{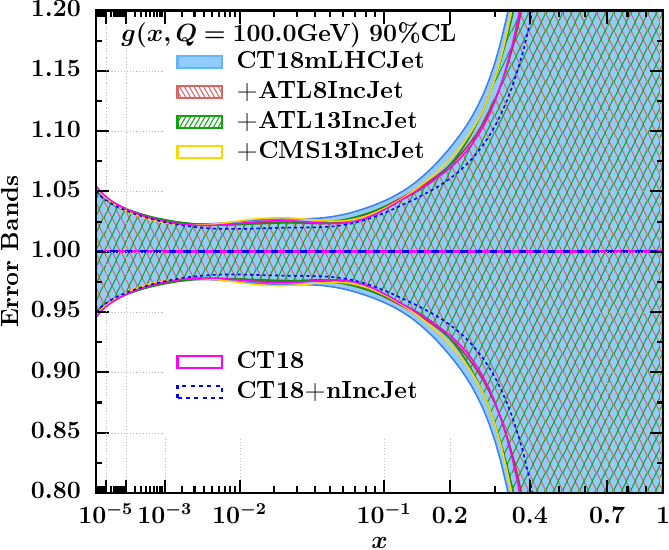}
	\includegraphics[width=0.49\linewidth]{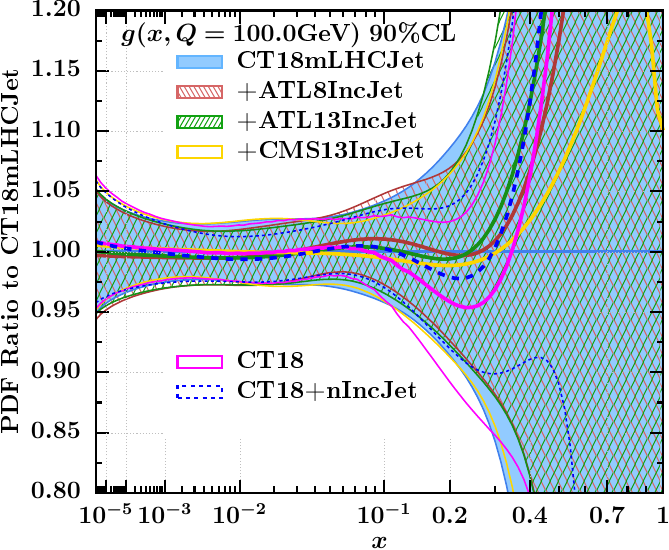}
	\caption{Comparison between the CT18mLHCJet, CT18, and CT18+nIncJet PDFs, and the newly obtained gluon PDFs. The left panel shows the error bands at 90\% C.L., while the right panel displays the ratios to the CT18mLHCJet baseline. 
 }\label{Fig:CT18mLHCJetvsIncJetUpdates}
\end{figure}

\begin{figure}[htbp]

\includegraphics[width=0.49\linewidth]{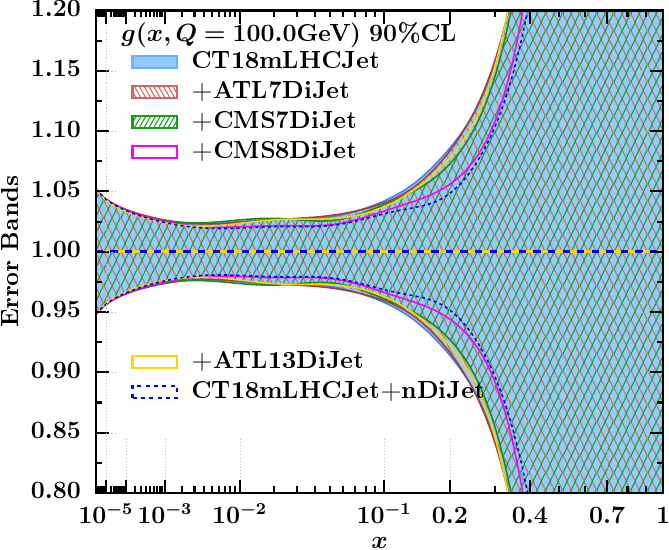}
\includegraphics[width=0.49\linewidth]{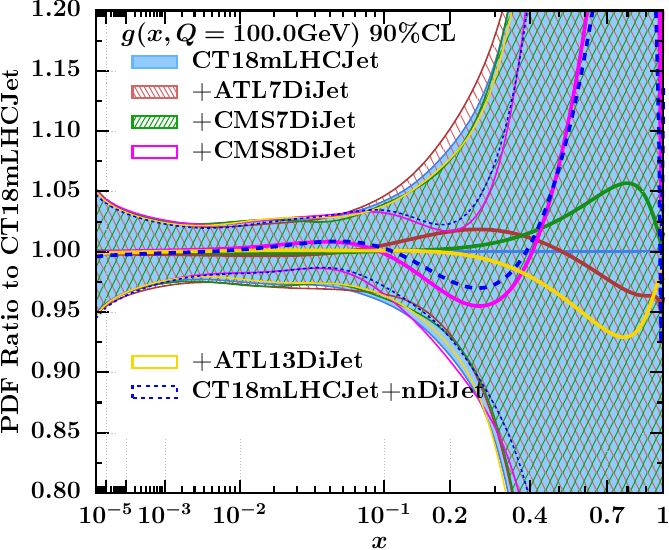}
\caption{Similar to Fig.~\ref{Fig:CT18mLHCJetvsIncJetUpdates}, but for dijet data sets.}\label{Fig:CT18mLHCJet+NewDiJet}
\end{figure}

The impact on the gluon PDF from individual inclusive jet and dijet data sets are compared in Fig.~\ref{Fig:CT18mLHCJetvsIncJetUpdates}
and Fig.~\ref{Fig:CT18mLHCJet+NewDiJet}, respectively.
All datasets favor a softer gluon distribution at a momentum fraction of $x \approx 0.3$, compared to the CT18mLHCJet baseline in Fig.~\ref{Fig:CT18mLHCJetvsIncJetUpdates}. 
Specifically, the ATLAS data at 7 TeV (ATL7IncJet) has the most significant impact, indicating a preference for a softer gluon distribution by approximately 7\%. This is followed by the CMS data at 7 TeV (CMS7IncJet), which shows a 5\% preference. The trend continues to with CMS data at 8 TeV (CMS8IncJet), indicating a 3\% preference, and CMS data at 13 TeV (CMS13IncJet), with a 1\% preference. In contrast, the ATLAS data at 13 TeV and 8 TeV show less significant deviations compared to the baseline.
In short, at $x \approx 0.3$, all datasets show a preference for a softer gluon distribution. The impact on the best-fit gluon PDF values is strongest to weakest as follows: ATL7IncJet (7\%) $>$ CMS7IncJet (5\%) $>$ CMS8IncJet (3\%) $>$ CMS13IncJet (1\%) $>$ ATL13IncJet $>$ ATL8IncJet.
At $x \approx 0.1$, the ATL8IncJet dataset provides the most constraining power compared to the inclusive jet datasets from ATLAS at 7 and 13 TeV, as well as the CMS datasets at 7, 8, and 13 TeV.

In Fig.~\ref{Fig:CT18mLHCJet+NewDiJet}, we present comparisons between the CT18mLHCJet gluon PDF and the gluon PDF obtained by including the dijet datasets (ATL7DiJet, CMS7DiJet, and CMS8DiJet) individually. We also include comparisons with the CT18mLHCJet+nDiJet gluon PDFs for convenience. 
The CMS8DiJet data suggests a preference for a softer gluon, while the ATL7DiJet data favors a harder gluon at a momentum fraction of x = 0.3, in comparison to the CT18mLHCJet baseline. 
In contrast, the CMS7DiJet data shows only a small deviation in the gluon PDF compared to the baseline.
In short, at $x \approx 0.3$, three dijet datasets show different preferences for the gluon distribution. The impact on the gluon uncertainty is strongest to weakest as follows: CMS8DiJet $>$ CMS7DiJet $>$ ATL7DiJet for  $x > 0.1$.

\subsection{Implication on the Drell-Yan production}
\begin{figure}
    \centering
    \includegraphics[width=0.49\linewidth]{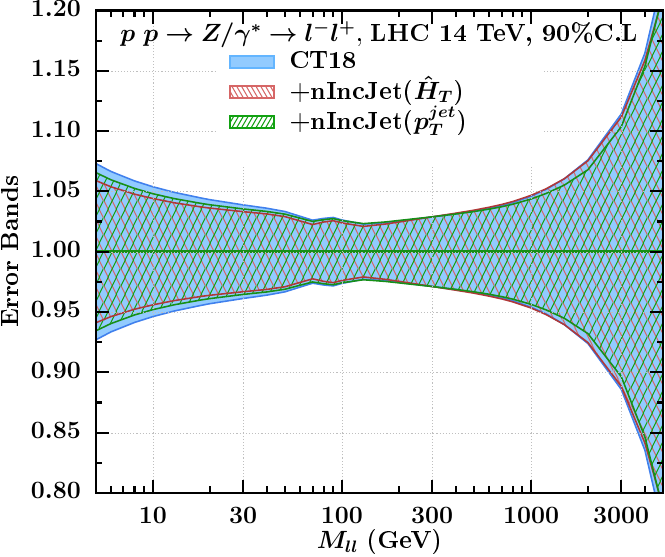}
    \includegraphics[width=0.49\linewidth]{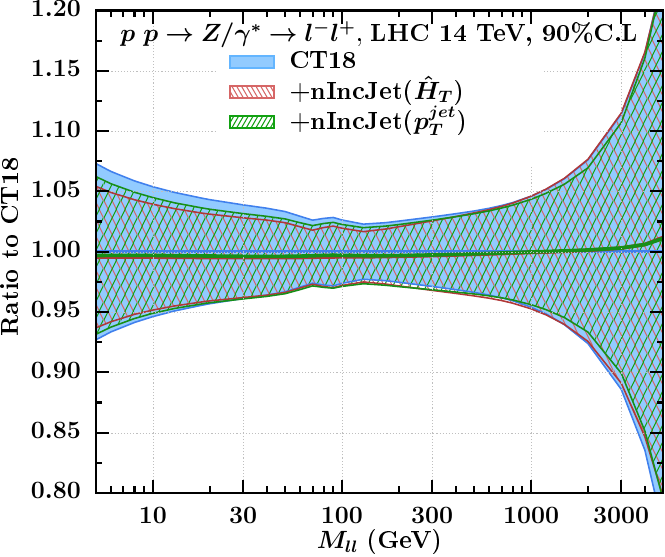}    
    \includegraphics[width=0.49\linewidth]{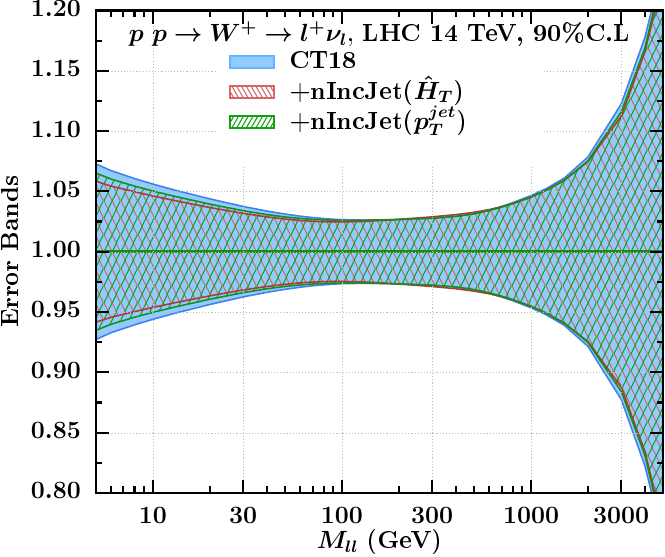}
\includegraphics[width=0.49\linewidth]{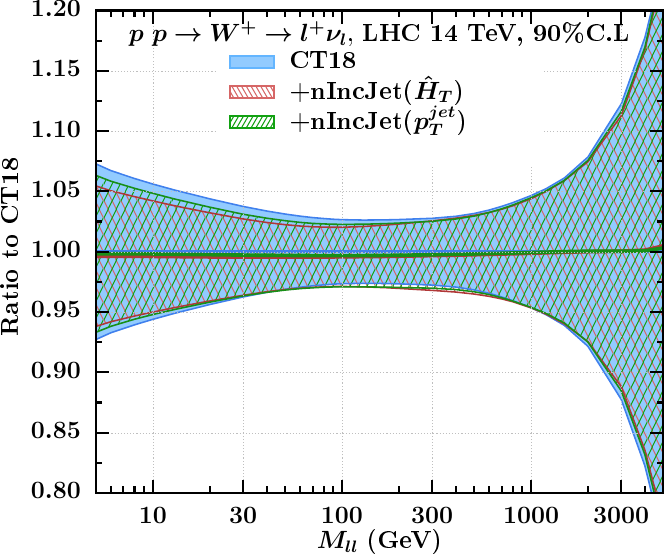}    
    \includegraphics[width=0.49\linewidth]{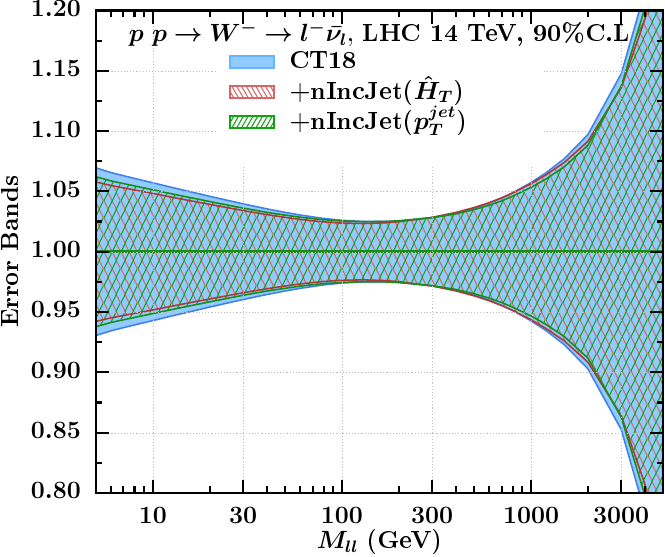}
\includegraphics[width=0.49\linewidth]{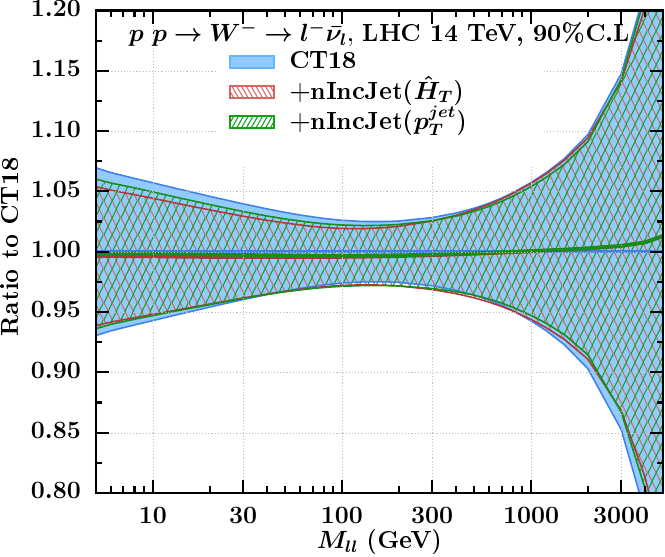}    
    \caption{
Similar to Fig. \ref{Fig:N3LO-Calc-H}, this figure illustrates the production of $Z$ (upper panel), $W^+$ (middle panel), and $W^-$ (lower panel).   
} \label{Fig:N3LO-Calc-WZ}
\end{figure}

Similar to the Higgs scenario in Fig.~\ref{Fig:N3LO-Calc-H}, here we present the N3LO predictions for the Drell-Yan production cross section $\dd\sigma/\dd M_{ll}^2$ in Fig.~\ref{Fig:N3LO-Calc-WZ}. As before, the theoretical calculations are performed with \texttt{n3loxs}~\cite{Baglio:2022wzu}, with the lepton-pair invariant mass $M_{ll}$ varying from 5 GeV to 5 TeV. 
In comparison with Fig.~\ref{Fig:N3LO-Calc-H}, the reduction in the PDF error band is less pronounced, particularly in the large mass region.
Additionally, the central predictions remain largely unchanged, indicating that the impact of the inclusive jet datasets on the quark parton distribution is relatively small.
A noticeable reduction in PDF uncertainty is observed in the small mass region, which can be attributed to the balance between the quark and gluon PDFs enforced by the momentum sum rule.
In comparison with the impact of the precision Drell-Yan data studied in Refs.~\cite{Sitiwaldi:2023jjp,Ablat:2024muy}, we observe the complementarity between these two types of datasets.

\subsection{$L_2$ Sensitivity with $T^2=10$}\label{L2SenT210}

\begin{figure}[!h]
    \includegraphics[width=\linewidth]{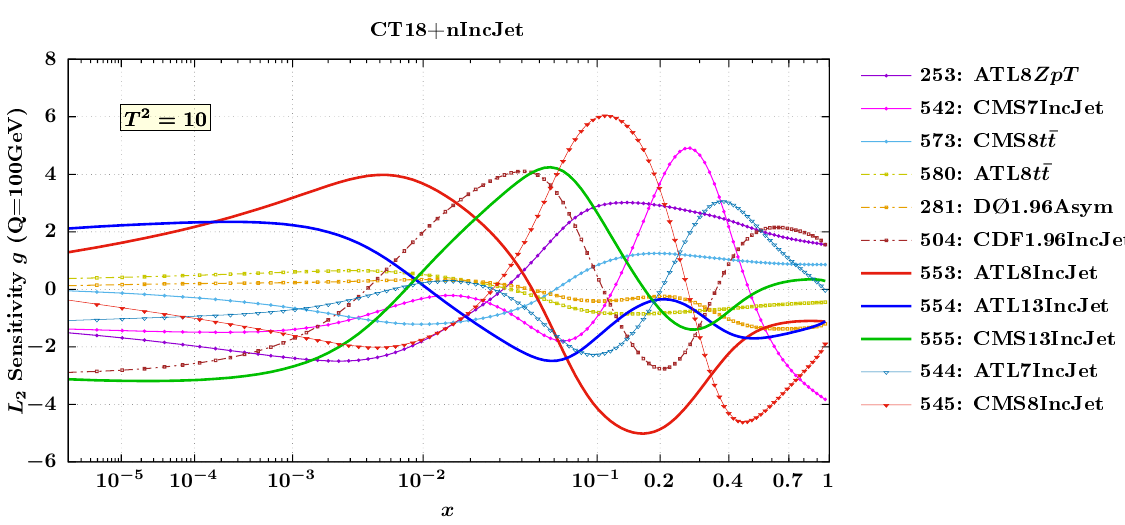}
    \caption{Similar to Fig.~\ref{Fig:L2Sensitivity}, but for $T^2=10$.}\label{Fig:L2SensitivityT210}
\end{figure}

$L_2$ Sensitivity is a way of viewing the pulls of the experiments used in a global PDF fit, for a particular parton flavor as a function $x$. For completeness, we provide the $L_2$ sensitivity 
plot in Fig.\ref{Fig:L2SensitivityT210} for the gluon PDF,  in CT18+nIncJet with $T^2=10$, which is the choice of the tolerance value adopted in Ref.~\cite{Jing:2023isu} to compare with the results from other global analysis groups, such as MRST20 and ATLAS. It provides similar information as Fig.~\ref{Fig:L2Sensitivity}, except the difference in its vertical scale, {\it i.e.,} the $L_2$ values.

\bibliographystyle{utphys}
\bibliography{Ref}

\end{document}